\documentclass[showpacs,preprintnumbers,amsmath,amssymb,twocolumn,10pt]{revtex4-1}
\usepackage{changes}
\usepackage{graphicx}
\usepackage{dcolumn}
\usepackage{bm}
\usepackage{subfigure}
\usepackage{float}
\usepackage{multirow}
\usepackage{booktabs}
\usepackage{amsmath}
\usepackage{cases}
\usepackage{latexsym,amssymb}
\usepackage[mathscr]{eucal}
\usepackage[switch*]{lineno}
\usepackage[bookmarks=true,
   colorlinks=true,
   linkcolor=blue,
   urlcolor=blue,
   citecolor=blue,
   bookmarks=true,
   hyperindex=true
]{hyperref}
\newcommand{\db}{\frac{\sin{(\delta_b b)}}{\delta_b}}
\newcommand{\dc}{\frac{\sin{(\delta_c c)}}{\delta_c}}

\begin{document}
\allowdisplaybreaks[4]
\title{Symmetric black-to-white hole solutions with a cosmological constant}

\author{Zhong-wen Feng\textsuperscript{1}}
\altaffiliation{Email: zwfengphy@cwnu.edu.cn}
\author{Qingquan Jiang\textsuperscript{1}}
\altaffiliation{Email: qqjiangphys@yeah.net}
\author{Yi Ling \textsuperscript{2,3,1}}
\altaffiliation{Email: lingy@ihep.ac.cn}
\author{Xiaoning Wu\textsuperscript{4,5}}
\altaffiliation{Email: wuxn@amss.ac.cn}
\author{Zhangping Yu\textsuperscript{1}}
\altaffiliation{Email: yuzp@ihep.ac.cn}
\vskip 0.5cm
\affiliation{1 School of Physics and Astronomy, China West Normal University, Nanchong 637009, China
\\ 2 Institute of High Energy Physics, Chinese Academy of Sciences, Beijing, 100049,  China
\\ 3 School of Physics, University of Chinese Academy of Sciences, Beijing 100049,  China
\\ 4 Institute of Mathematics, Academy of Mathematics and System Science and Hua Loo-Keng Key Laboratory of Mathematics, \\
Chinese Academy of Sciences, Beijing 100190, China
\\ 5 School of Mathematical Sciences, University of Chinese Academy of Sciences,  Beijing 100190, China}

\begin{abstract}
For a system with a Hamiltonian constraint, we demonstrate that its dynamics is invariant under different choices of the lapse function, regardless of whether the Hamiltonian incorporates quantum corrections. Applying this observation to the interior of black-to-white holes, we analyze its dynamics with different choices of the lapse function. The results explicitly show that the leading-order expansion of both metrics  proposed by Rovelli et al. (Class. Quant. Grav. \textbf{35}, 225003 (2018); Class. Quant. Grav. \textbf{35}, 215010 (2018)) and Ashtekar et al. (Phys. Rev. Lett. \textbf{121}, 241301 (2018); Phys. Rev. D \textbf{98}, 126003 (2018))  exhibit identical behavior near the transition surface. Therefore, in this sense the black-to-white hole model proposed by Rovelli et al., (Class. Quant. Grav. \textbf{35}, 225003 (2018); Class. Quant. Grav. \textbf{35}, 215010 (2018)) may be interpreted as a coarse-grained version of the solution within the framework of loop quantum gravity. The black-to-white hole solutions with exact symmetry between the black hole and white hole regions are constructed by appropriately fixing the quantum parameters in the effective theory of loop quantum gravity. This approach circumvents the issue of amplification of mass, which could arise from a mass difference between the black hole and white hole, and provides a way to link the solutions obtained by minisuperspace quantization to those in the covariant approach. Finally, the black-to-white hole solutions with a cosmological constant are constructed. The numerical solutions for the interior of the black-to-white hole with a cosmological constant are obtained, and their symmetric behavior is also discussed.
\end{abstract}

\maketitle
\section{Introduction}
\label{intro}
The successful detection of gravitational waves emanating from black holes~\cite{LIGOScientific:2016aoc} and the observation of the shadow of black holes ~\cite{LIGOScientific:2019fpa,EventHorizonTelescope:2020qrl} have announced the coming of a new age for black hole physics.  As a theory that predicts the existence of black holes, the general relativity (GR) plays a pivotal role in scrutinizing their properties including the structure of spacetime, causality as well as the thermodynamic properties. Nonetheless, as one delves deeper into the evolution of a massive star into a black hole, a quandary emerges---the gravitational collapse inexorably culminates and a singularity unavoidably appears at the center of a black hole, wherein all matter converges to a region with infinite curvature~\cite{Penrose:1964wq,ch5-}, which signals the ultraviolet incompleteness of classical GR as a theory for the description of the spacetime of a black hole.  The emergence of a singularity also implies that existing physical theories fail under such extreme conditions, leading to a significant gap in our understanding on the physical processes inside the horizon of a black hole. It is widely believed that as the curvature of spacetime approaches the Planckian scale, the quantum effects of gravity would become significant enough to change the geometry of spacetime. Therefore, quantum gravity (QG) would offer a promising way for us to come out from the enigma of the singularity of spacetime \cite{Witten:1991yr,Rovelli:1997qj,Bojowald:2005qw,Saini:2014qpa,Ashtekar:2021kfp,Barcelo:2015noa}. Before a complete theory of quantum gravity is established, people have attempted to construct black holes without singularity (which are called as non-singular black holes or regular black holes) at both the phenomenological level and the semi-classical level. At the phenomenological level, one may construct the metric for regular black holes by introducing the exotic matter fields into the standard Einstein field equations, and the presence of such exotic matter fields might ultimately ascribe to the effects of QG. Historically, the famous Bardeen black hole \cite{Bardeen1968} and Hayward black hole  \cite{Hayward:2005gi} are prominent examples in this route. Recently, some novel solutions for regular black holes have also been constructed in this manner  \cite{Xiang:2013sza,Balart:2014jia,Fan:2016hvf,Li:2016yfd,Nojiri:2017kex,He:2017ujy,Rodrigues:2018bdc,deSousaSilva:2018kkt,Simpson:2018tsi,Zhang:2019acn,Melgarejo:2020mso,Burzilla:2020utr,Bonanno:2020fgp,Estrada:2020tbz,Torres:2022twv,Lan:2023cvz,Junior:2023ixh,Mazza:2023iwv,Luongo:2023aib}. In particular, a black-to-white hole solution characterized by the sub-Planckian curvature is constructed in Refs.~\cite{Bianchi:2018mml,DAmbrosio:2018wgv}. The key observation leading to a sub-Planckian curvature is that the parameter $l$, which could be understood as the regulator reflecting the scale that the quantum corrections of gravity begin to have an impact, may not have the same magnitude as the fundamental scale-Planck length, but depending on the mass of the black hole as well. As a result, one finds that the maximal value of the Kretschmann scalar curvature is not only finite, but also bounded from above by the Planck energy density. Inspired  by this work, some novel regular black holes with sub-Planckian curvature have been investigated in Refs.~\cite{Rovelli:2018cbg,Brahma:2018cgr,Carballo-Rubio:2018jzw,Bodendorfer:2019jay,Alesci:2019pbs,BenAchour:2020gon,Ling:2021olm,LingLingYi:2021rfn,Rignon-Bret:2021jch,Ling:2022vrv,Zeng:2022yrm,Zeng:2023fqy,Feng:2023pfq}. Nevertheless, we intend to point out that all the metrics of regular black holes proposed at the phenomenological level assume the existence of unknown exotic matter fields. Usually, one just derives the form of their energy stress tensor by requiring that  the Einstein equations should be satisfied, while their Lagrangian in terms of such matter fields is unknown. Essentially, one is urged to provide the theoretical foundation for such regular black holes by solving the equations of motion with matter fields as fundamental variables. It is worthwhile to point out that recently some typical regular black holes have been derived in the framework of gravity coupled to nonlinear  electrodynamics~\cite{Ayon-Beato:1998hmi,Dehghani:2019xhm,Ayon-Beato:1999kuh,Bronnikov:2000vy,Dymnikova:2004zc,Balart:2014cga,Allahyari:2019jqz,Guerrero:2020uhn,Kruglov:2021stm,Kokoska:2021lrn,Ali:2022zox,Capozziello:2023vvr}, and other modified gravity theories, for instance, see Refs.~\cite{Moffat:2014aja,Modesto:2016max,Simpson:2019cer,Carballo-Rubio:2019fnb,Carballo-Rubio:2021wjq,Bakopoulos:2021liw,Zeng:2021kyb,Li:2023yyw,Nian:2023xmr,Lin:2023jqz}. On the contrary, the semi-classical approach is devoted to construct the effective equations of motion at first, which may receive significant quantum corrections of gravity due to quantum geometry and thus exhibit distinct behavior from that in the classical theory of GR, and then try to derive the black-to-white metric as the solution to these modified equations. Usually, such metrics may look more complicated, but their theoretical foundation is more robust. Recently, such effective equations of motion for Schwarzschild black hole have been constructed in the framework of loop quantum gravity (LQG) \cite{Ashtekar:2018lag,Ashtekar:2018cay}. Previously, the relevant works on applying LQG techniques to the singularity problem of black hole can be found in literatures, for instance in Refs.~\cite{Nicolai:2005mc,Bojowald:2005cb,Cartin:2006yv,Ashtekar:2005qt,Modesto:2005zm,Boehmer:2007ket,Chiou:2008nm,Brannlund:2008iw,Reuter:2012id,Dadhich:2015ora,Corichi:2015xia,Olmedo:2017lvt,Cortez:2017alh,Perez:2017cmj,Yonika:2017qgo,Joe:2014tca,Afrin:2022ztr}. In Refs.~\cite{Ashtekar:2018lag,Ashtekar:2018cay}, one significant  improvement is  incorporating new quantum parameters into the semi-classical Hamiltonian constraint. These parameters are set to be Dirac observables, which means that along each dynamic trajectory they are constants, thereby enabling the simultaneous depiction of the influence of LQG on the Schwarzschild black hole. Particularly, it is interesting enough to notice that the quantum parameters exhibit the same mass-dependent behavior as that in Refs.~\cite{Bianchi:2018mml,DAmbrosio:2018wgv}. In this new effective theory, singularity is eliminated and the black hole may undergo a transition to a white hole, presenting a novel avenue for addressing the singularity problem of black holes, and some subsequent works can be found in Refs.~\cite{Martin-Dussaud:2019wqc,Kelly:2020uwj,Zhang:2020qxw,Zhang:2021wex,Zhang:2023yps,Geiller:2020xze,Daghigh:2020fmw,Ashtekar:2020ifw,Hong:2022thd,ElizagaNavascues:2022npm,Lewandowski:2022zce,Han:2023wxg,Zhang:2023okw,Fu:2023drp,Gan:2022oiy,Gan:2024rga}.

The motivations of the present work are following. Firstly, we intend to reveal some connections between two black-to-white solutions in Refs.~\cite{Bianchi:2018mml,DAmbrosio:2018wgv} and Refs.~\cite{Ashtekar:2018lag,Ashtekar:2018cay}. On one hand, the metric form of the black-to-white solution in Refs.~\cite{Bianchi:2018mml,DAmbrosio:2018wgv} are quite simple and elegant, but proposed at the phenomenological level. In addition, its metric form  only applies to the interior of the horizon, and if one tries to extend it to the region outside the horizon, one would find that the energy conditions would be violated everywhere, as pointed out in  Ref.~\cite{Feng:2023pfq}. Here our effort may provide a theoretical foundation for this solution and embed it into the effective theory of black holes in the context of LQG. On the other hand, the metric form of the black-to-white solution in Refs.~\cite{Ashtekar:2018lag,Ashtekar:2018cay} is obtained by explicitly solving the effective equations of motion which are derived in the context of LQG, thus with a robust theoretical foundation. Nevertheless, we notice that in comparison with Refs.~\cite{Bianchi:2018mml,DAmbrosio:2018wgv} a different lapse function is considered in Refs.~\cite{Ashtekar:2018lag,Ashtekar:2018cay}, which is very helpful for one to write down the analytical expression of solutions. However, in this coordinate system, the location of the  transition surface does not occur at the original singularity with $T_{cl} = - \infty$, but at some place with finite $T_{q}$. Although it is understood that the time coordinate $T_{cl}$ in classical theory is different from that $T_q$ in semi-classical theory, this scenario perhaps is not quite intuitive as that in Ref.~\cite{Bianchi:2018mml}, where the transition surface from black hole to white hole occurs exactly at the location of the original singularity, namely at $T_{cl}\approx T_q$. In other words, the singularity is replaced by a transition from a black hole to a white hole. The scenario depicted in Ref.~\cite{Bianchi:2018mml} is manifestly similar to that in loop quantum cosmology (LQC), where the cosmological singularity is replaced by a bounce from an attracting  universe to an expanding universe \cite{Bojowald:2001xe,Ashtekar:2003hd,Ashtekar:2004eh,Thiemann:2006cf,Thiemann:2001gmi,Thiemann:2002nj}. Furthermore, the values of the quantum parameters considered in Refs.~\cite{Ashtekar:2018lag,Ashtekar:2018cay} transition surface  at finite $T$
always result in spacetime asymmetry between the black hole region and the white hole region, which would lead to the problem of amplification of mass \cite{Ashtekar:2018cay,Corichi:2015xia}.  This issue is initially reflected in the metric of Ref.~\cite{Corichi:2015xia}, where for a a solar black
hole, the mass of the white hole will be increased by a factor $10^{114}$. This increase in mass would imply an unexplained energy source, contradicting the fundamental conservation laws of physics. Moreover, if the spacetime is maximally extended, the mass of the white hole would increase indefinitely, resulting in an unphysical divergence and instability in the extended spacetime. Although this issue has  been improved in Ref.~\cite{Ashtekar:2018cay}, where the amplification of mass in Schwarzschild case is tiny and thus can be ignored, since this amplification of mass will be influenced by the cosmological constant and will be enhanced with the extension of spacetime, whether it is  always small enough in general cases is not justified yet. Besides, an open problem remaining in Refs.~\cite{Ashtekar:2018lag,Ashtekar:2018cay} is that they ignore the issue of covariance because the minisuperspace quantization does not show how the temporal and spatial variations of fields are related. The metric in Refs.~\cite{Ashtekar:2018lag,Ashtekar:2018cay} is obtained starting with a symmetry reduced theory, and it is not known if there is a 4-dimensional covariant action whose symmetry reduction would yield the dynamical equations. Additionally, the asymmetry between the black hole region and the white hole region also seems to imply that the model is not consistent with covariant model because the effective metric obtained in a covariant formalism always exhibits symmetric behavior between the black hole and white hole, as demonstrated in Ref.~\cite{zhang2024blackholescovarianceeffective}. Inspired by the above works, the following questions can be raised: Is it possible to provide a more intuitive picture for the black hole to white hole transition where the original singularity is replaced by the transition surface? Could one construct a black-to-white hole solution with spacetime symmetry between the black hole and white hole regions such that one need not worry about the amplification of mass even in the general cases? Is it possible to demonstrate the consistency between the minisuperspace model and the covariant model? All of the above questions will be answered in the affirmative in this work.

Subsequently, we also intend to apply this universal framework to other types of black holes in a spacetime with a cosmological constant.  In light of the above motivations, in this paper we will firstly show that the dynamics of the geometry inside the horizon is always independent of the choice of the lapse function no matter if  the quantum correction is taken into account. This observation links the black-to-white solution in Refs.~\cite{Bianchi:2018mml,DAmbrosio:2018wgv} to the one in Refs.~\cite{Ashtekar:2018lag,Ashtekar:2018cay}. The leading order expansion of the former solution near the transition surface is identical to that of the latter solution with the same lapse function. Then, we focus on the symmetry between the black hole and the white hole. It is found that the effective metric may exhibit a symmetric behavior when the quantum parameters are  appropriately fixed such  that the issue trouble of mass amplification can be avoided. Finally, we extend the above discussion to construct black-to-white hole solutions with a cosmological constant. The numerical solutions for the interior of black-to-white hole with a cosmological constant are obtained, and the symmetric behavior of the metric between the black hole region and the white hole region is also discussed.

The paper is organized as follows. In section~\ref{sec2}, we show  that the dynamics of a system with Hamiltonian constraint is equivalent under the  different choices of  the lapse function, no matter if this system is classical or semi-classical. Then, we apply this to discuss the dynamics of the interior of the Schwarzschild black hole and the corresponding black-to-white hole with quantum corrections. We will focus on the symmetry between the black hole region and the white hole region with different lapse functions. In section~\ref{sec3}, we extend this discussion  to construct black-to-white hole solutions with a cosmological constant, showing the effect of the cosmological constant on the geometry and analyzing the spacetime symmetry. In particular, we propose an iteration method to obtain the values of quantum parameters in the presence of a cosmological constant to guarantee that the area enclosed by two plaquettes on the transition surface remains the area gap. The paper ends with conclusions and discussions in section~\ref{sec4}.

To simplify the notation, in this work we adopt the Planck units $G=k_B=c=\hbar=1$.

\section{The dynamics of the interior region of Schwarzschild black hole}
\label{sec2}
\subsection{The classical dynamics in canonical formalism}
\label{sec2-1}
In this section, we will demonstrate that for a  system with a Hamiltonian constraint, one is free to choose the lapse function to investigate its dynamics since the system is invariant under the re-parameterization of the time coordinate.

It is well known that for a system with a Hamiltonian constraint, one is free to choose the lapse function, which corresponds to the specification of the time coordinate. To see this, consider two different lapse functions $N_1$ and $N_2$ such that the resultant Hamiltonians are $H(N_1)=N_1 h$ and $H(N_2)=N_2 h$, respectively, with the Hamiltonian constraint $h \approx 0$. The corresponding time coordinates are denoted as $T_1$  and $T_2$, respectively. The re-parameterization invariance requires that the spacetime geometry is described by the same metric, namely ${\text{d}}{s^2} =  - N_1^2{\text{d}}T_1^2 +  \cdots  =  - N_2^2{\text{d}}T_2^2 + \cdots$.

Obviously, the above relation holds if and only if the two time coordinates are transformed as follows
\begin{align}
\frac{{{\text{d}}{T_1}}}{{{\text{d}}{T_2}}} = \frac{{{N_2}}}{{{N_1}}}.
\label{eq:time_trans}
\end{align}
We remark that here we have assumed that the cross terms $g_{TX}$ in the metric are zero, namely $g_{TX}=0$. With this condition, it is straightforward to show that the canonical equations of motion with lapse function $N_1$ are equivalent to those with lapse function  $N_2$. Specifically, taking the canonical equation for the variable $b$ (which will appear as the variable for the Schwarzschild black hole in the next) as an example, one has
\begin{align}
\label{eq:eom_of_b_a}
\frac{{\text{d}} b}{{\text{d}} T_1}=  \frac{\partial \left(N_1 h \right)}{\partial p_b} =  \left( N_1 \frac{\partial h}{\partial p_b}+h\frac{\partial N_1}{\partial p_b}\right) \approx  N_1 \frac{\partial h}{\partial p_b} .
\end{align}
On the other hand, by applying the relation in Eq.~(\ref{eq:time_trans}), one can directly derive the canonical equation with $T_2$ as the time coordinate as follows
\begin{align}
\label{eq:eom_of_b_b}
\frac{{\text{d}} b}{{\text{d}} T_2} \!=  \!\frac{{\text{d}} T_1}{{\text{d}} T_2} \frac{{\text{d}} b}{{\text{d}} T_1}  \!=   \!\frac{N_2}{N_1} \frac{{\text{d}} b}{{\text{d}} T_1}  \!\approx   \!\frac{N_2}{N_1}  N_1 \frac{\partial h}{\partial p_b} \! =   \! N_2 \frac{\partial h}{\partial p_b}  \approx  \frac{\partial \left(N_2 h \right)}{\partial p_b} .
\end{align}
The second equality follows from Eq.~(\ref{eq:time_trans}), the third weak equality comes from Eq.~(\ref{eq:eom_of_b_a}), and the last weak equality results from $h \approx 0$. Therefore, if the solution to Eq.~(\ref{eq:eom_of_b_a}) is $b \left(T_1 \right)$, then $b \left(T_2 \right):=b\left(T_1 \left(T_2 \right)\right)$ constitutes the solution to Eq.~(\ref{eq:eom_of_b_b}), where the time transformation $T_1 \left(T_2 \right)$ is given by Eq.~(\ref{eq:time_trans}). The equations of motion, as well as the solutions for other canonical variables, can be derived analogously. Thus, it is concluded that for a system with Hamiltonian constraint, one is free to choose the lapse function $N$, which corresponds to choose a specific time coordinate for the system. However, at root all the dynamics with different lapse functions are  equivalent. In particular, we stress that the above argument is applicable to all the systems with Hamiltonian constraint, including the effective theory of black holes after the quantum correction is taken into account, where just a different form of Hamiltonian is introduced.

Next, we briefly review the canonical formalism of gravity vacuum in terms of the Ashtekar variables for a spacetime with spherical symmetry closely following the scheme presented in Refs.~\cite{Ashtekar:2018lag,Ashtekar:2018cay}. In a homogeneous but non-isotropic Kantowski-Sachs spacetime,  the connection as the configuration variable  and the spatial triad as the conjugate variable are reduced to the following form~\cite{Ashtekar:1987gu}
\begin{equation}
\label{eq2}
A_a^i{\tau _i}{\text{d}}{x^a} = \bar c{\tau _3}{\text{d}}x \!+\! \bar b{r_o}{\tau _2}{\text{d}}\theta - \bar b{r_o}{\tau _1}\!\sin \theta {\text{d}}\phi + {\tau _3}\cos \theta {\text{d}}\phi,
\end{equation}
\begin{equation}
\label{eq3}
E_i^a{\tau ^i}{\partial _a}  = {{\bar p}_c}r_o^2{\tau _3}\sin \theta {\partial _x} +{{\bar p}_b}{r_o}{\tau _2}\sin \theta {\partial _\theta }- {{\bar p}_b}{r_o}{\tau _1}{\partial _\phi, }
\end{equation}
where the triad variables are defined as $\bar c = {c \mathord{\left/
 {\vphantom {c {{L_o}}}} \right. \kern-\nulldelimiterspace} {{L_o}}}$, $\bar b = {b \mathord{\left/ {\vphantom {b {{r_o}{\text{ }}}}} \right. \kern-\nulldelimiterspace} {{r_o}{\text{ }}}}$, ${{\bar p}_c} = {{{p_c}} \mathord{\left/ {\vphantom {{{p_c}} {r_o^2}}} \right. \kern-\nulldelimiterspace} {r_o^2}}$, and ${{\bar p}_b} = {{{p_b}} \mathord{\left/ {\vphantom {{{p_b}} {{L_o}{r_o}}}} \right. \kern-\nulldelimiterspace} {{L_o}{r_o}}}$. ${L_o}$ is the length of the fiducial cell and ${r_o}$ is the radius in the fiducial metric, while $\tau_i$ are SU(2) generators related to the Pauli spin matrices $\sigma_i$.  As a result, the phase space is described by two pairs of conjugate variables $\left( {c,{p_c}} \right)$ and $\left( {b,{p_b}} \right)$. The  corresponding Poisson brackets are given by $\left\{ {c,{p_c}} \right\} = 2\gamma$ and $\left\{ {b,{p_b}} \right\} = \gamma$ with  the Barbero-Immirzi parameter $\gamma$~\cite{Ashtekar:2018lag,Ashtekar:2018cay}.

The interior of the spacetime inside the horizon  is homogeneous such that the metric is only time-dependent, which takes the form as
\begin{equation}
\label{eq4}
{\text{d}}{s^2} =  - N ^2 {\text{d}}{T ^2} + \frac{{p_b^2}}{{\left| {{p_c}} \right|L_o^2}}{\text{d}}{x^2} + \left| {{p_c}} \right| {{\text{d}}{\Omega ^2}},
\end{equation}
where ${\text{d}}{\Omega ^2} = {\text{d}}{\theta ^2} + {\sin ^2}\theta {\text{d}}{\phi ^2}$ represents the line element of the spherical surface, and $T$ is the time coordinate associated with the lapse function $N$. Plugging the form of the metric into the action, one can derive the Hamiltonian formalism with a constraint. According to Refs.~\cite{Ashtekar:2005qt,Modesto:2005zm}, the classical Hamiltonian constraint is
\begin{align}
\label{eq6}
{\mathcal{H} =  - \frac{N}{{2 {\gamma ^2}}}\frac{{\operatorname{sgn} \left( {{p_c}} \right)}}{{\sqrt {\left| {{p_c}} \right|} }}\left[ {2bc{p_c} + \left( {{b^2} + {\gamma ^2}} \right){p_b}} \right]}.
\end{align}

In order to derive the canonical equations,  one specifies  the lapse function as~\cite{Ashtekar:2005qt,Modesto:2005zm}
\begin{align}
\label{eqa1}
N \left(T \right) = \frac{{\gamma \operatorname{sgn} \left( {{p_c}} \right){{\left| {{p_c}(T )} \right|}^{\frac{1}{2}}}}}{b},
\end{align}
and denotes the corresponding time variable as $T$. By substituting Eq.~(\ref{eqa1}) into Eq.~(\ref{eq6}), the Hamiltonian reads \cite{Corichi:2015xia}
\begin{align}
\label{eqa2}
\mathcal{H}\left( N \left(T\right) \right) =  - \frac{1}{{2\gamma }}\left[ {2c{p_c} + \left( {b + \frac{{{\gamma ^2}}}{b}} \right){p_b}} \right],
\end{align}
which leads to the equations of motion as follows
\begin{subequations}
\label{eqa3}
\begin{align}
&\dot b  = \left\{{b,\mathcal{H}} \right\} = \gamma \frac{{\partial \mathcal{H}}}{{\partial {p_b}}} =  - \frac{1}{{2b}}\left( {{b^2} + {\gamma ^2}} \right),  \label{eqa3-a}\\
&\dot c  = \left\{ {c,\mathcal{H}} \right\} = 2\gamma \frac{{\partial \mathcal{H}}}{{\partial {p_c}}} =  - 2c, \label{eqa3-b}\\
&{{\dot p}_b}  = \left\{ {{p_b},\mathcal{H}} \right\} =  - \gamma \frac{{\partial \mathcal{H}}}{{\partial b}} = \frac{{{p_b}}}{{2{b^2}}}\left( {{b^2} - {\gamma ^2}} \right), \label{eqa3-c}\\
&{{\dot p}_c}  = \left\{ {{p_c},\mathcal{H}} \right\} =  - 2\gamma \frac{{\partial \mathcal{H}}}{{\partial c}} = 2{p_c}, \label{eqa3-d}
\end{align}
\end{subequations}
where the ``dot" denotes the time derivative with respect to $T$. By solving the dynamical equations above with the Hamiltonian constraint $H\left( N \right) \approx 0$ and properly fixing the integration constants, one obtains the solutions as
\begin{subequations}
\label{eqa4}
\begin{align}
&{p_c}\left( T \right)  = 4{m^2}{e^{2T}},
 \\
&c\left( T \right)  = \frac{{\gamma {L_o}}}{{4m}}{e^{ - 2T}},\\
& {p_b}\left( T \right)  =  - 2m{L_o}{e^T}{\left( {{e^{ - T}} - 1} \right)^{\frac{1}{2}}},
\\
& b\left( T \right)  = \gamma {\left( {{e^{ - T}} - 1} \right)^{\frac{1}{2}}},
\end{align}
\end{subequations}
which gives rise to the metric for the interior of the Schwarzschild black hole as follows
\begin{align}
\label{eqa5}
{\text{d}}{s^2} &=  - \frac{{4{m^2}{e^{2T}}}}{{{e^{ - T}} - 1}}{\text{d}}{T^2} + \left( {{e^{ - T}} - 1} \right){\text{d}}{x^2}+ {\left( {4{m^2}{e^{2T}}} \right)^{\text{2}}}{\text{d}}{\Omega ^2}.
\end{align}
In this coordinate system, the event horizon of the Schwarzschild black hole is located at $T=0$, and the singularity occurs at $T= -\infty $, which indicates that the covering region of metric~(\ref{eqa5}) is $\left( { - \infty ,0} \right)$.

Next, we intend to link the metric form adopted in Refs.~\cite{Bianchi:2018mml,DAmbrosio:2018wgv}  to the above metric by means of an explicit coordinate transformation.  Based on the above discussion, we choose a new lapse function as follows
\begin{align}
\label{eq6+}
N'\left(\tau \right) = \frac{{2\gamma \operatorname{sgn} \left( {{p_c}} \right){{\left| {{p_c}} \right|}^{\frac{1}{4}}}}}{b},
\end{align}
and the corresponding time coordinate is denoted as $\tau$. By substituting Eq.~(\ref{eq6+}) into Eq.~(\ref{eq6}), the corresponding Hamiltonian reads
\begin{align}
\label{eq7}
{\mathcal{H} \left( N' \right) =  - \frac{1}{{\gamma }}\left[ {2p_c^{3/4}c + \frac{{{p_b}}}{{p_c^{1/4}}}\left( {b + \frac{{{\gamma ^2}}}{b}} \right)} \right]}.
\end{align}
The equations of motion are
\begin{subequations}
\label{eq8}
\begin{align}
&\dot b\left(\tau \right)   =- \frac{1}{p_c^{1/4} b}\left( {{b^2} + {\gamma ^2}} \right),  \label{eq8-a}\\
&\dot c\left(\tau \right)   = -4 p_c^{-1/4} c\\
&{{\dot p}_b\left(\tau \right)}  = \frac{{{p_b}}}{{p_c^{1/4}} b^2}\left( {b^2 - \gamma^2} \right),  \label{eq8-c}\\
&{{\dot p}_c\left(\tau \right)}  = 4p_c^{3/4}, \label{eq8-d}
\end{align}
\end{subequations}
where the ``dot" denotes the time derivative with respect to $\tau$ and $H \approx 0$ is used to simplify the equations.  The time transformation is given by
\begin{align}
\frac{{{\text{d}}{\tau} }}{{{\text{d}}T}} = \frac{N}{{N'}} = \frac{{p_c^{{1 \mathord{\left/
 {\vphantom {1 4}} \right.
 \kern-\nulldelimiterspace} 4}}}}{2} = \sqrt {\frac{m}{2}} {e^{\frac{T}{2}}},
\end{align}
and one gets ${\tau} = \sqrt{2m} e^{\frac{T}{2}}$ and $T = 2\ln \left( {{{\tau}  \mathord{\left/
 {\vphantom {\tau  {\sqrt {2m} }}} \right.
 \kern-\nulldelimiterspace} {\sqrt {2m} }}} \right)$. The integration constant is determined by requiring the singularity to be located at ${\tau}=0$. The solutions to Eq.~(\ref{eq8}) are then
 \begin{subequations}
 \label{eq10}
\begin{align}
&b\left({\tau} \right) =  \frac{{\gamma \sqrt {2m - {{\tau} ^2}} }}{{\tau} }, \\
&c\left({\tau} \right) =  -\frac{{{L_o}m\gamma }}{{{{\tau} ^4}}}, \\
&{p_b}\left({\tau} \right) = {L_o}{\tau} \sqrt {2m - {{\tau} ^2}}, \\
&{p_c}\left({\tau} \right) = {{\tau} ^4}.
\end{align}
\end{subequations}
Notably, ${p_c} c$ is a Dirac observable, so that one can define ${p_c} c = {L_o m} \gamma$. Now, plugging Eq.~(\ref{eq10}) into Eq.~(\ref{eq4}), one finds the metric is expressed as
\begin{align}
\label{eq11}
{\text{d}}{s^2} =  - \frac{{4{\tau ^4}}}{{2m - {\tau ^2}}}{\text{d}}{\tau ^2} + \frac{{2m - {\tau ^2}}}{{{\tau ^2}}}{\text{d}}{x^2} + {\tau ^4} {\text{d}}{\Omega^2},
\end{align}
which is nothing but the metric form for the interior of the Schwarzschild black hole presented in Refs.~\cite{Bianchi:2018mml,DAmbrosio:2018wgv}. Now the event horizon is located at  ${\tau} = \sqrt {2m}$ and the singularity occurs at ${\tau} = 0$. The original region of the interior of Schwarzschild black hole is $0 < {\tau}< \sqrt {2m}$.  It is observed in Refs.~\cite{Bianchi:2018mml,DAmbrosio:2018wgv} that if one introduces some quantum parameter $l$ into the metric to eliminate the singularity, then  one may extend the time coordinate of metric~(\ref{eq11}) to $-\sqrt{2m} < \tau < \sqrt{2m}$, leading to a white hole region
\begin{align}
\label{eq:Rovelli_RBH_ds2}
\mathrm{d} s^{2}= -\frac{4\left({\tau}^{2}+l\right)^{2}}{2 m-\tau^{2}} \!\mathrm{d} {\tau}^{2}\!+\!\frac{2 m-{\tau}^{2}}{{\tau}^{2}+l}\! \mathrm{d} x^{2}\!+\!\left({\tau}^{2}\!+\!l\right)^{2}\! \mathrm{d} \Omega^{2},
\end{align}
where $l \sim {m^{1/3}}$ is a constant depending on the mass of the black hole. It is noted that this black-to-white hole exhibits an exact symmetric behavior between the black hole region and the white hole region since the metric is invariant under the transformation $\tau\rightarrow -\tau$.  Conversely, it is not manifest to extend spacetime to a white hole region simply by introducing some quantum parameter into the metric in Eq.~(\ref{eqa5}) and then by time extension. The above discussion indicates that the choice of the lapse function does not alter the geometry of spacetime (a similar conclusion is also presented in Ref.~\cite{Ongole:2023pbs}), but it might constrain the possible extension of spacetime coordinate after some quantum corrections are taken into account. In the next subsection, we first review the effective black-to-white hole solution for the interior based on the Hamiltonian~(\ref{eq6}) and  the lapse function~(\ref{eqa1}), and then investigate its dynamics by changing the lapse function to Eq.~(\ref{eq6+}).

\subsection{Effective dynamics of the interior Schwarzschild black-to-white hole}
\label{sec2-2}
The holonomy correction plays a core role in LQG and LQC. The key point is that the local variables which are defined at a specific point of spacetime should be replaced by non-local variables that are defined over a small region of spacetime, such as the holonomy of the connection.  This replacement may change the evolution picture of the universe, as well as the black hole dramatically. Following Refs.~\cite{Ashtekar:2018lag,Ashtekar:2018cay}, we first consider the lapse function~(\ref{eqa1}) with the time coordinate $T$. By implementing the replacement $b \to {{\sin \left( {{\delta _b}b} \right)} \mathord{\left/ {\vphantom {{\sin \left( {{\delta _b}b} \right)} {{\delta _b}}}} \right. \kern-\nulldelimiterspace} {{\delta _b}}}$ and $c \to {{\sin \left( {{\delta _c}c} \right)} \mathord{\left/ {\vphantom {{\sin \left( {{\delta _c}c} \right)} {{\delta _c}}}} \right. \kern-\nulldelimiterspace} {{\delta _c}}}$ \cite{Ashtekar:2005qt,Boehmer:2007ket}, where the quantum parameters $\delta_b$ and $\delta_c$ reflect the QG effect with holonomy corrections. The effective lapse function is expressed as
\begin{align}
{N_{\text{eff}} \left(T \right) = \frac{{\gamma \operatorname{sgn} \left( {{p_c}} \right){\delta _b}{{\left| {{p_c}(T )} \right|}^{\frac{1}{2}}}}}{{\sin \left( {{\delta _b}b} \right)}}},
\end{align}
where we denote the corresponding time variable by $T$ for simplicity, but it should be kept in mind that  the time coordinate is  now different from that in classical theory. The effective Hamiltonian is then formulated as
\begin{align}
\label{eq13}
{\mathcal{H}_ \text{eff} \left(N_\text{eff} \right)\! = \! - \frac{1}{{2\gamma }} \!
 \left[ {2\frac{{\sin \left( {{\delta _c}c} \right)}}{{{\delta _c}}} p_c \! + \!p_b \! \left(\! {\frac{{\sin \left( {{\delta _b}b} \right)}}{{{\delta _b}}}\!\! +\!\! \frac{{{\gamma ^2}{\delta _b}}}{{\sin \left( {{\delta _b}b} \right)}}} \!\right)} \right]}.
\end{align}
It is evident that for ${{\delta _b}\to 0}$ and ${{\delta _c}\to 0}$,  the classical lapse function and the Hamiltonian are restored. However, for non-vanishing ${\delta _b}$ and ${\delta _c}$, both the lapse function and the Hamiltonian are distinct from the classical cases (i.e., Eq.~(\ref{eq6}) and Eq.~(\ref{eqa1})); thus, a different dynamical behavior is expected. As a matter of fact, the equations of motion for the connection and triad components are derived as follows:
\begin{subequations}
\label{eq14}
\begin{align}
&\dot b  = -\frac{1}{2}\left(\db + \frac{\gamma^2 \delta_b}{\sin(\delta_b b)} \right),\\
&\dot c  = -2 \dc, \\
&{{\dot p}_b}  = \frac{p_b}{2}\cos(\delta_b b)\left(1-\frac{\gamma^2\delta_b^2}{\sin^2(\delta_b b)} \right),\\
&{{\dot p}_c}  = 2 p_c\cos \left( {{\delta _c}c} \right).
\end{align}
\end{subequations}
The solutions are
\begin{subequations}
\label{eq:sol_of_eff_eom_N1}
\begin{align}
&\cos \left(\delta_b b(T) \right) =b_0  \tanh \left[\frac{1}{2} \left( b_0 T + 2 \tanh^{-1} \left( \frac{1}{b_0}\! \right) \right) \right],\label{eq:sol_of_eff_eom_N1_b}\\
&\tan\left(\frac{\delta_c c(T)}{2}\right) = \mp \frac{\gamma L_o \delta_c}{8 m} e^{-2T},\label{eq:sol_of_eff_eom_N1_c}\\
&p_b(T)=-2m\gamma L_o\frac{\delta_b\sin \left(\delta_b b \left(T \right) \right)}{\sin^2(\delta_b b(T)) + \gamma^2 \delta_b^2},\label{eq:sol_of_eff_eom_N1_pb}\\
&p_c \left(T \right)=4m^2\left(e^{2T}+\frac{\gamma^2 L_o^2\delta_c^2}{64 m^2} e^{-2T} \right)\label{eq:sol_of_eff_eom_N1_pc},
\end{align}
\end{subequations}
where $b_0=\sqrt{1+\gamma^2 \delta_b^2}$. From Eq.~(\ref{eq:sol_of_eff_eom_N1_pc}), it is found that  $p_c$ takes a nonzero minimum value at the special time $T_\mathcal{T}$
\begin{align}
{l^2}: & = {\left. {{p_c}} \right|_{{\text{min}}}} = {p_c}\left( {{T_\mathcal{T}}} \right) = {L_o}\gamma m{\delta _c},\label{eq:l} \\
T_\mathcal{T} &= \frac{1}{2}\ln\left(\frac{l^2}{8m^2}\right). \label{eq:T_T}
\end{align}
The space-like hypersurface at $T=T_\mathcal{T}$ is denoted by $\mathcal{T}$. The analysis in Refs.~\cite{Ashtekar:2018lag,Ashtekar:2018cay} demonstrates that this hypersurface is a transition surface that separates the trapped region and anti-trapped region. In the classical limit $\delta_b$, $\delta_c \rightarrow 0$, one has $T_\mathcal{T} \rightarrow -\infty$, corresponding to the singularity of the classical black hole. In this sense, the transition surface $\mathcal{T}$ supplants the classical singularity in the quantum corrected geometry. The interior region of the classical black hole, which corresponds to $-\infty<T<0$, now maps to the region $T_\mathcal{T}<T<T_{\text{BH}}$, with $T_{\text{BH}}=0$ being the location of the black hole horizon. In the absence of the singularity,  it can now be extended to a white hole region, corresponding to $T_{\text{WH}}<T<T_\mathcal{T}$, where $T_{\text{WH}}=-(4/b_0)\tanh^{-1}(1/b_0)$ denotes the location of the white hole horizon.

Next, let us consider the dynamics with the lapse function in Eq.~(\ref{eq6+}). Once the quantum correction is taken into account, the effective lapse function and effective Hamiltonian become
\begin{align}
N'_{\text{eff}} \left({\tau} \right)  = \frac{{2\gamma \operatorname{sgn} \left( {{p_c}} \right){\delta _b}{{\left| {{p_c} \left({\tau} \right)} \right|}^{\frac{1}{4}}}}}{{\sin \left( {{\delta _b}b} \right)}},
\end{align}
\begin{align}
&\mathcal{H}_\text{eff} \left(N'_\text{eff} \right)
\nonumber \\
&=  - \frac{1}{{\gamma }}
\left[ {2p_c^{3/4}\frac{{\sin \left( {{\delta _c}c} \right)}}{{{\delta _c}}} + \frac{{{p_b}}}{{p_c^{1/4}}}\left( {\frac{{\sin \left( {{\delta _b}b} \right)}}{{{\delta _b}}} + \frac{{{\gamma ^2}{\delta _b}}}{{\sin \left( {{\delta _b}b} \right)}}} \right)} \right].
\end{align}
Here the corresponding time variable is still denoted by ${\tau}$. The new equations of motion now read
\begin{subequations}
\label{eq:eff_eom_N2}
\begin{align}
&\dot b\left({\tau} \right)  =-\frac{1}{p_c^{1/4}}\left(\db + \frac{\gamma^2 \delta_b}{\sin(\delta_b b)} \right), \\
&\dot c\left({\tau} \right)   = -4 p_c^{-1/4} \dc, \\
&{{\dot p}_b\left({\tau} \right)}  = p_c^{-1/4}p_b \cos(\delta_b b)\left(1-\frac{\gamma^2\delta_b^2}{\sin^2(\delta_b b)} \right),\\
&{{\dot p}_c\left({\tau} \right)}  = 4 p_c^{3/4}\cos \left( {{\delta _c}c} \right).
\end{align}
\end{subequations}
Two time coordinates $T$ and $\tau$ transform as follows
\begin{align}
\frac{{\text{d}} \tau}{{\text{d}} T} \! =\! \frac{N_{\text{eff}}}{N'_{\text{eff}}}\!=\!\frac{p_c^{1/4}}{2}\!=\!\sqrt{\frac{m}{2}}\left(e^{2T}\!+\!\frac{\gamma^2 L_o^2\delta_c^2}{64 m^2} e^{-2T} \right)^{\!\!1/4}\!,
\end{align}
\begin{align}
\tau \left(T \right)& =\frac{\Gamma \left(\frac{7}{8} \right)}{\Gamma \left(\frac{3}{8} \right)}\sqrt{\pi l}-\frac{\left(64 m^2 e^{2T}+ \frac{l^4}{m^2}e^{-2 T}\right)^{1/4}}{2\left(1+\frac{64m^4}{l^4} e^{4 T}\right)^{1/4}}
\nonumber\\
&\times {}_2F_1 \left(-\frac{1}{4}, -\frac{1}{8}; \frac{7}{8}; \frac{64m^4}{l^4}e^{4 T} \right) \label{eq:time_trans_sol},
\end{align}
where $l$ is given by Eq.~(\ref{eq:l}) and ${}_2F_1$ is the hypergeometric function. The integration constant is fixed by requiring the transition surface to be located  at $\tau_{ \mathcal{T}}=0$. Then, the solutions to Eq.~(\ref{eq:eff_eom_N2}) are given by Eq.~(\ref{eq:sol_of_eff_eom_N1}) composed with time transformation $T \left(\tau \right)$, where $T \left(\tau \right)$ is the inverse function of Eq.~(\ref{eq:time_trans_sol}). The horizons of black hole and white hole are now located at ${\tau _{{\text{BH}}}} = {\tau \left( {{T_{{\text{BH}}}}} \right)}$ and ${\tau _{\text{WH}}} = {\tau \left( {{T_{{\text{WH}}}}} \right)}$, respectively.

\begin{figure}[htbp]
\centering
\includegraphics[width=0.38 \textwidth]{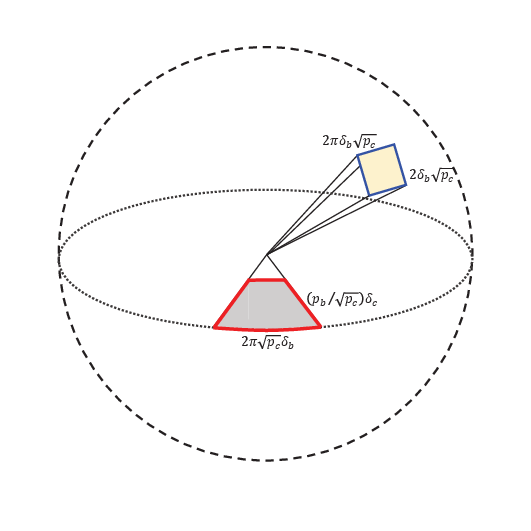}
\caption{Schematic diagram of the physical area of two plaquettes. The area of gray plaquette is ${\delta _b}{\delta _c}\left( {{{\left. {2\pi \left| {{p_b}} \right|} \right|}_\mathcal{T}}} \right)$ on the transition surface, while the area of orange plaquette is ${\left( {{\delta _b}} \right)^2}\left( {{{\left. {4\pi {p_c}} \right|}_\mathcal{T}}} \right)$ on the transition surface.}
\label{figure1}
\end{figure}

Following Refs.~\cite{Ashtekar:2018lag,Ashtekar:2018cay}, the quantum parameter $\delta_{b}$ is considered to be the length of each link constituting the plaquette  within the $\theta-\phi$ 2-sphere, while $\delta_{c}$  is the fractional length of the links in the $x$-direction within the plaquette in the $\theta-x$ and $\phi-x$ planes in a fiducial cell (see Fig.~\ref{figure1} for a schematic representation). By requiring that the area enclosed by these  two plaquettes on the transition surface $\mathcal{T}$ is equal to the area gap $\Delta$:
\begin{align}
\label{eqx1}
{\left. {2\pi {\delta _c}{\delta _b}\left| {{p_b}} \right|} \right|_\mathcal{T}} = \Delta  ,\quad {{\left. {4\pi \delta _b^2{p_c}} \right|}_\mathcal{T}} = \Delta.
\end{align}
The quantum parameters $\delta_{b}$ and $\delta_{c}$ can be constrained in the large $m$ limit as follows:
\begin{align}
\label{eq:Ashtekar_delta_bc}
\delta_{b}=\left(\frac{\sqrt{\Delta}}{\sqrt{2 \pi} \gamma^{2} m}\right)^{1 / 3}, \quad L_{o} \delta_{c}=\frac{1}{2}\left(\frac{\gamma \Delta^{2}}{4 \pi^{2} m}\right)^{1/3}.
\end{align}

\begin{figure*}[htbp]
\setlength{\belowcaptionskip}{1cm}
\centering
\includegraphics[scale=0.7]{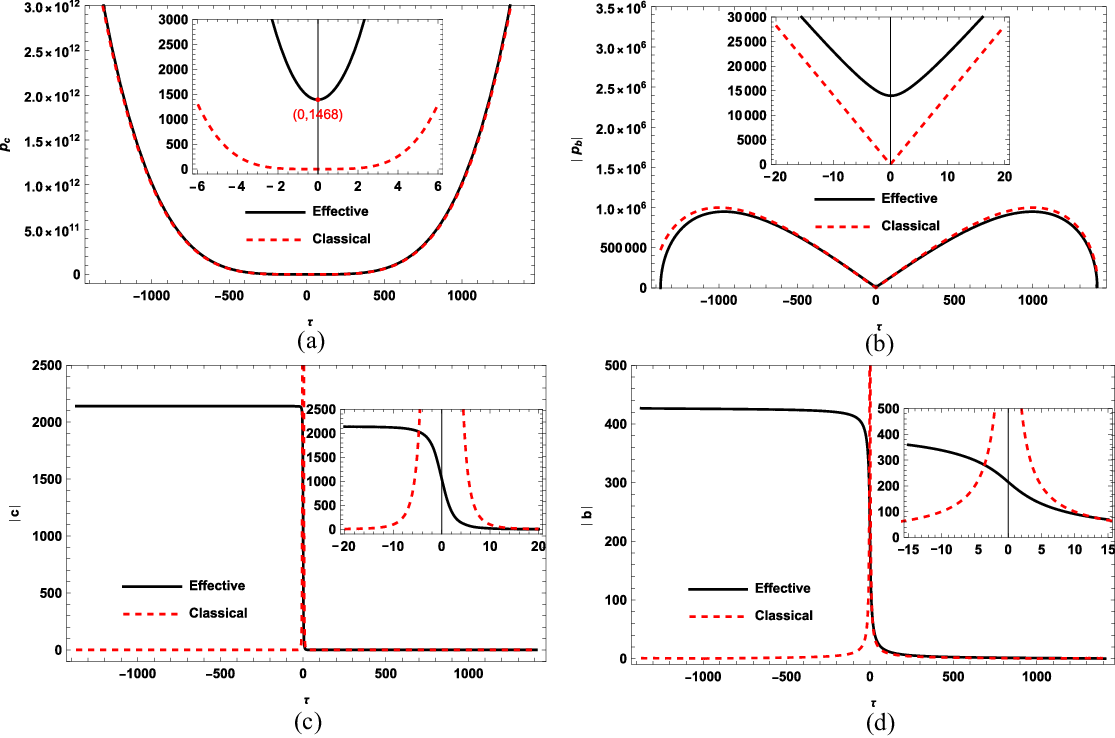}
\caption{Time ($\tau$) evolution of $p_c$, $p_b$, $c$ and $b$ in the Schwarzschild interior for $m=10^6$, $L_o=\Delta=\gamma=1$,  ${\delta _b} = {\left( {\sqrt \Delta  /\sqrt {2\pi } {\gamma ^2}m{\text{ }}} \right)^{1/3}}$ and ${L_o}{\delta _c} = {\left( {\gamma {\Delta ^2}/4{\pi ^2}m{\text{ }}} \right)^{1/3}}/2$.  (a) $p_c$ versus $\tau$. (b) $\left|{p_b} \right|$ versus $\tau$. (c) $\left| c \right|$ versus $\tau$. (d) $\left| b \right|$ versus $\tau$. The black solid curve represents the effective Schwarzschild black-to-white hole case, while the red dashed curve denotes the classical Schwarzschild black hole case. }
\label{figure2}
\end{figure*}

It is noteworthy that both quantum parameters are proportional to $m^{-1/3}$, whereas then $l$ goes as $m^{1/3}$. Now we consider the dynamics of the interior with respect to the time coordinate $\tau$. The time evolution of $p_c$, $p_b$, $c$ and $b$  in the interior of the black-to-white hole are illustrated in Fig.~\ref{figure2}.

In Fig.~\ref{figure2}, the red dashed curves represent the dynamics of the metric components and their conjugate variables of the classical Schwarzschild black hole. It should be noted that all these curves terminate at $\tau = 0$ due to the existence of the singularity, and the rest can be obtained by the time evolution from  $-\tau$  to $0$ and from $\tau$ to $0$. The black solid curves describe the evolution of the metric components for a quantum corrected Schwarzschild black-to-white hole, which evolves in time from $\tau_\text{BH}$ to $\tau_\text{WH}$. It is evident that these curves are continuous and the metric components do not vanish at $\tau = 0$, indicating that the singularity is erased by QG effect, which opens a window for matter passing from the black hole region to the white hole region.

\subsection{The symmetry of the geometry between black hole and white hole}
\label{sec2-3}
From the perspective of symmetry,  Fig.~\ref{figure2} shows that $p_c\left(\tau \right)$ and $p_b\left(\tau\right)$ are approximately even functions, whereas $c\left(\tau\right)$ and $b\left(\tau\right)$ are approximately odd functions. Therefore, there exists an approximate symmetry between the black hole region  and the white hole region. However, this symmetry is not exact. In fact, when the quantum parameters $\delta_b$ and $\delta_c$ are set as Eq.~(\ref{eq:Ashtekar_delta_bc}), one can derive the ratio of the  black hole mass to the white hole mass as
\begin{align}
\frac{m_{\text{WH}}}{m_{\text{BH}}} = 1+\mathcal{O}\left[\left(\frac{\ell_{\text{Pl}}}{m}\right)^{\frac{2}{3}}\ln\left(\frac{m}{\ell_{\text{Pl}}}\right)\right],
\end{align}
which implies that the masses of the black hole and white hole are equal only in the limit of large $m$.

We intend to propose a different scheme to fix quantum parameters to guarantee that the metric of the black-to-white hole exhibits an exact symmetric behavior between the black hole region and the white hole region. As a result, the mass of the black hole is exactly equal to that of the white hole. Utilizing Eq.~(\ref{eq:sol_of_eff_eom_N1}), it is observed that
\begin{align}
p_c \left(T_{\text{BH}} \right) & -p_c \left(T_{\text{WH}} \right)
\nonumber \\
&=\frac{1}{16m^2}\left(e^{-2 T_{\text{WH}}}-1\right)\left(64m^4 e^{2 T_{\text{WH}}}-l^4\right).
\end{align}
Then, by setting $p_c \left(T_{\text{BH}} \right)=p_c \left(T_{\text{WH}} \right)$, one obtains that $8m^2 e^{T_{\text{WH}}}=l^2$, or
\begin{align}
\label{eq:delta_bc_relation}
\exp \left[ { - \frac{4}{{{b_0}}}{{\tanh }^{ - 1}}\left( {\frac{1}{{{b_0}}}} \right)} \right] = \frac{{{L_o}\gamma {\delta _c}}}{{8m}},
\end{align}
where $b_0=\sqrt{1+\gamma^2 \delta_b^2}$. Note that this condition can also be written as $\cos\left(\delta_b b \left(T_\mathcal{T} \right)\right) = 0$ or $T_{\text{WH}} = 2 T_{\mathcal{T}}$. Therefore, if we set $\delta_{b}$ as before, but change the setup of ${L_o}{\delta _c}$ as described above, then the quantum parameters become
\begin{align}
\label{eq:delta_bc}
\delta_{b}=\left(\frac{\sqrt{\Delta}}{\sqrt{2 \pi} \gamma^{2} m}\right)^{1 / 3},
\end{align}
\begin{align}
\label{eq:delta_bc-1}
{L_o}{\delta _c} = \frac{{8m}}{\gamma }\exp \left[ { - \frac{4}{{{b_0}}}{{\tanh }^{ - 1}}\left( {\frac{1}{{{b_0}}}} \right)} \right].
\end{align}
Now, one finds that the radius of the black hole horizon $r_\text{BH}=\sqrt{p_c \left(T_{\text{BH}} \right)}$ is equal to the radius of the white hole horizon $r_{\text{WH}}=\sqrt{p_c \left(T_{\text{WH}} \right)}$. Obviously,  the leading order of ${L_o}{\delta _c}$ is consistent with Eq.~(\ref{eq:Ashtekar_delta_bc}), namely ${L_o}{\delta _c} = \frac{1}{2}{\left( {{{\gamma {\Delta ^2}} \mathord{\left/ {\vphantom {{\gamma {\Delta ^2}} {4{\pi ^2}m}}} \right.
 \kern-\nulldelimiterspace} {4{\pi ^2}m}}} \right)^{1/3}} + \mathcal{O}\left( {{m^{ - 4/3}}} \right)$. One can treat the set in Eq.~(\ref{eq:delta_bc}) as the modification of Eq.~(\ref{eq:Ashtekar_delta_bc}). Furthermore, by virtue of the solutions presented in Eq.~(\ref{eq:sol_of_eff_eom_N1}),  one can directly verify that $p_b(T)$ and $p_c(T)$ are even functions with respect to  $T=T_\mathcal{T}$, while $b(T)$and $c(T)$ are odd functions with respect to  $T=T_\mathcal{T}$. In fact, we have
\begin{align}
c \left( {{T_\mathcal{T}} + T} \right) &+ c\left( {{T_\mathcal{T}} - T} \right)  = \frac{2}{{{\delta _c}}} \left[ {{{\tan }^{ - 1}}\!\left( { \mp {e^{{T_{{\text{WH}}}}}}{e^{ - {T_{{\text{WH}}}} - 2T}}} \right)} \right. \nonumber \\
&\left. { + {{\tan }^{ - 1}}\left( { \mp {e^{{T_{{\text{WH}}}}}}{e^{ - {T_{{\text{WH}}}} + 2T}}} \right)} \right] =  \frac{2}{{{\delta _c}}}\left( {n + \frac{\pi }{2}} \right).
\end{align}
The symmetry of $b\left(T \right)$ and $p_b(T)$ can be similarly verified. Additionally, one can confirm that the time transformation ${\tau}\left(T\right)$ is an odd function with respect to $T=T_\mathcal{T}$. Consequently, $p_b\left({\tau}\right)$ and $p_c\left({\tau}\right)$ are even functions with respect to ${\tau}=0$, while $b\left({\tau}\right)$ and $c\left({\tau}\right)$ are odd functions with respect to ${\tau}=0$. Thus, the metric exhibits symmetry under the transformation ${\tau} \rightarrow -{\tau}$ as well. We will prove this symmetry again in a different manner in section \ref{sec3-3}, where a cosmological constant is involved.

\begin{figure}[H]
\centering
\includegraphics[scale=0.8]{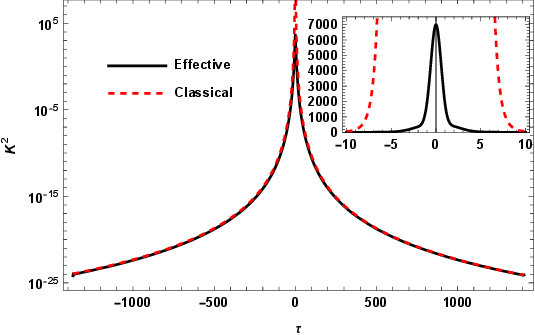}
\caption{Time ($\tau$) evolution of the Kretschmann scalar in the Schwarzschild interior for $m=10^6$, ${\delta _b} = {\left( {\sqrt \Delta  /\sqrt {2\pi } {\gamma ^2}m} \right)^{1/3}}$ and ${L_o}{\delta _c} = \frac{{8m}}{\gamma }\exp \left\{ { - \left[ {4{{\tanh }^{ - 1}}\left( {1/{b_0}} \right)} \right]/{b_0}} \right\}$. The black solid curve represents the effective Schwarzschild black-to-white hole case, while the red dashed curve denotes the classical Schwarzschild black hole case.}
\label{figure3}
\end{figure}

Thanks to the symmetry of the metric, all curvature scalars are even functions and attain their extremal values at ${\tau}=0$. Recall that $\mathcal{T}$ corresponds to the singularity of  classical black hole, one expects that ${\tau} =0$ should be a point with the maximal value. One can justify this through numerical analysis, as demonstrated in Fig.~\ref{figure3}, where the Kretschmann scalar has a maximal value at $\mathcal {T}$.

Next, we intend to embed the metric proposed in Refs.~\cite{Bianchi:2018mml,DAmbrosio:2018wgv} into the framework of LQG by showing that it can be viewed as the approximation solution of the above equations. Furthermore, we will show that these two metrics are consistent with the covariant metric proposed in Ref.~\cite{zhang2024blackholescovarianceeffective} under the same approximation. Let us examine the behavior of the metric near the transition surface. The components of the metric can be expressed as
\begin{align}
-N'^2_{\text{eff}}\left({\tau} \right)  \!= \! -4l \left(b_0^2-1 \right)-4 \left(b_0^2-1  \right)\left(4+ b_0^4 \right){{\tau}}^2\! +\!  \mathcal{O}\left({{\tau}}^4 \right),
\end{align}
\begin{align}
\frac{{p_b \left({{\tau}} \right)^2}}{{\left| {{p_c  \left({\tau} \right)}} \right|L_o^2}}  & =\frac{4 \left(b_0^2-1 \right)m^2}{b_0^4 l^2}
\nonumber\\
& -\frac{4 \left(b_0^2-1 \right) \left(b_0^4-2b_0^2+8 \right)m^2}{b_0^4l^3}{\tau}^2+ \mathcal{O} \left({{\tau}}^4 \right),
\end{align}
\begin{align}
p_c \left({\tau} \right) &=l^2+8l{{\tau}}^2+\mathcal{O} \left({{\tau}}^6 \right),
\end{align}
where $b_0$ and $l$ satisfy condition (\ref{eq:delta_bc_relation}). Note that $l\sim m^{1/3}$, $b_0\sim1$ and $(b_0^2-1)=\gamma^2\delta_b^2\sim m^{-2/3}$, the effective metric can be written as
\begin{align}
\label{eq:Sch_eff_ds2}
\text{d}s^2& = - N'^2_{\text{eff}}\left({\tau} \right) \text{d}{{\tau}}^2+\frac{{p_b({{\tau}})^2}}{{\left| {{p_c({{\tau}})}} \right|L_o^2}}{\text{d}}{x^2} + \left| {{p_c({{\tau}})}} \right| {{\text{d}}{\Omega ^2}}
\nonumber \\
& \sim \!-\left(m^{- \frac{1}{3}}\!+ \!m^{-\frac{2}{3}}{\tau}^2 \right){\text{d}} {{\tau}}^2 \!+ \! \left(m^{\frac{2}{3}}\! -\! m^{\frac{1}{3}}{{\tau}}^2 \right){\text{d}} x^2
\nonumber \\
&+ \left(m^{\frac{2}{3}}+m^{\frac{1}{3}}{{\tau}}^2 \right){\text{d}} \Omega^2 + \mathcal{O} \left({{\tau}}^4 \right).
\end{align}

On the other hand, the metric~(\ref{eq:Rovelli_RBH_ds2}) which is given in Refs.~\cite{Bianchi:2018mml,DAmbrosio:2018wgv} can also be written as
\begin{align}
\mathrm{d} s^{2}&=-\frac{4\left({{\tau}}^{2}+l\right)^{2}}{2 m-{{\tau}}^{2}} \mathrm{d} \tau^{2}+\frac{2 m-{{\tau}}^{2}}{\tau^{2}+l} \mathrm{~d} x^{2}+\left({{\tau}}^{2}+l\right)^{2} \mathrm{d} \Omega^{2}
\nonumber \\
&=  \left(-\frac{2l^2}{m} -\frac{l^2+4lm}{m^2}{{\tau}}^2  \right){\text{d}} {{\tau}}^2 +  \left(\frac{2m}{l}-\frac{l+2m}{l^2}{{\tau}}^2  \right){\text{d}} x^2
\nonumber \\
& +\left(l^2 + 2l {{\tau}}^2 \right){\text{d}} \Omega^2 + \mathcal{O} \left({{\tau}}^4 \right)
\nonumber \\
&\sim - \left(m^{-\frac{1}{3}}+m^{-\frac{2}{3}}{{\tau}}^2 \right){\text{d}}{{\tau}}^2 + \left(m^{\frac{2}{3}}-m^{\frac{1}{3}}{{\tau}}^2 \right){\text{d}} x^2
\nonumber \\
& + \left(m^{\frac{2}{3}}+m^{\frac{1}{3}}{{\tau}}^2 \right){\text{d}} \Omega^2 + \mathcal{O} \left({{\tau}}^4 \right).
\end{align}
Interestingly enough, a covariant model with the $\bar{\mu}$-scheme is also proposed in Ref.~\cite{zhang2024blackholescovarianceeffective}, and the metric is given by
\begin{align}
\mathrm{d} s^2=-f \mathrm{d}t^2+\mu^{-1} f^{-1}\mathrm{d} x^2+x^2 \mathrm{d}\Omega^2,
\label{eq:3}
\end{align}
where
\begin{align}
&f=1-\frac{2m}{x},
\nonumber \\
&\mu=1+\frac{\zeta^2}{x^2}
\left(1-\frac{2m}{x}\right),
\nonumber \\
&\zeta=\sqrt{4\sqrt{3}\pi \gamma^3 l_p^2},
\end{align}
and $l_p$ denotes the Planck length. This coordinate covers the black hole exterior region and the black hole interior region. To cover the entire black hole region and the white hole region, we introduce a new coordinate system $\left(T, X\right)$ which satisfies $x=R(T), t=X$. Then, metric~(\ref{eq:3}) can be written as
\begin{align}
\mathrm{d} s ^ { 2 } = - N^{2} \mathrm{d} T ^ { 2 } + \left( \frac {2m}{R} - 1 \right) \mathrm{d} X ^ { 2 } + R^{2} \mathrm{d} \Omega^{ 2 },
\end{align}
where
\begin{align}
R\left(T\right) + \frac { R ^ { 3 }\left(T\right) } { \zeta ^ { 2 } } \sin ^ { 2 } \left( T \right) = 2 m,
\nonumber \\
N\left(T\right) = \frac { 2 \zeta R ^ { 2 }\left(T\right) } { \zeta ^ { 2 } + 3 R ^ { 2 }\left(T\right) \sin ^ { 2 } \left( T \right) }.
\end{align}
It is clear that the metric is exactly symmetric between the black hole region and the white hole region, and the radius of the transition surface is $l=R \left(\pi/2 \right)\sim m^{1/3}$, which is consistent with our model. Next we examine the behavior of the metric near the transition surface. Without loss of generality, we set $\zeta$ to unity here. We further introduce a time transformation $T=\frac{\pi}{2}+m^{-\frac{1}{6}}\tau$, then the metric can be written as
\begin{align}
\mathrm{d}s^2=&- \left[ \frac{4l^4}{\left(3l^2+1 \right)^2m^{\frac{1}{3}}} +  \frac{8\left(-2l^4+3l^6+9l^8+4l^3m \right)}{\left(3l^2+1 \right)^4m^{\frac{2}{3}}}\tau^2  \right] \mathrm{d}\tau^2
\nonumber\\
&+ \left[\frac{2m}{l}-1+\frac{2\left(l-2m \right)m^{\frac{2}{3}}}{l^2 \left(3l^2+1 \right)}\tau^2 \right]\mathrm{d} X^2
\nonumber\\
&+ \left[l^2+\frac{2l\left(2m-l \right)}{3l^2+1}m^{-\frac{1}{3}}\tau^2 \right]\mathrm{d}\Omega^2+\mathcal{O}\left(\tau^4\right)
\nonumber\\
&\sim- \left(m^{-\frac{1}{3}}+ m^{-\frac{2}{3}} \tau^2\right)\mathrm{d}\tau^2+\left(m^{\frac{2}{3}}-m^{\frac{1}{3}}\tau^2 \! \right)\mathrm{d}X^2
\nonumber\\
&+ \left(m^{\frac{2}{3}}+m^{\frac{1}{3}}\tau^2 \right)\mathrm{d}\Omega^2.
\end{align}
It explicitly demonstrates that the metric in the covariant model exhibits the identical behavior near the transition surface as the  symmetric metric in our model. In summary, after comparing the behavior of all three metrics near the transition surface, we intend to conclude that the leading-order expansion of all the three metrics appeared in literature, namely the covariant metric in Ref.~\cite{zhang2024blackholescovarianceeffective}, the metric in Refs.~\cite{Bianchi:2018mml,DAmbrosio:2018wgv}  and the symmetric metric in our article, exhibits identical behavior near the transition surface, and can be considered consistent under the coarse-graining approximation.

Using the approximate metric~(\ref{eq:Sch_eff_ds2}), it is easy to show that the leading terms of the maximal value of the curvature scalars are mass independent:
\begin{align}
R^2\left|_{\mathcal{T}}\right.&=4 \left(1+m^{-\frac{2}{3}} \right)^2,\\    R_{ab}R^{ab}\left|_{\mathcal{T}}\right.&=2 \left(2+2m^{-\frac{2}{3}}+m^{-\frac{4}{3}} \right),\\
 K^2\left|_{\mathcal{T}}\right.&=4 \left(3+m^{-\frac{3}{4}} \right),\\
 C_{abcd}C^{abcd}\left|_{\mathcal{T}}\right.&=\frac{4}{3} \left(2-m^{-\frac{2}{3}} \right)^2.
\end{align}
This is a significant characteristic of quantum corrected black holes, suggesting that the curvature scalars are bounded at the Planck scale  \footnote{The discrepancy between the numerical results based on Eq.~(\ref{eq:Sch_eff_ds2}) and the above analytical results comes from the fact that many factors have been discarded in Eq.~(\ref{eq:Sch_eff_ds2}). Their contribution to  the curvature may not small, but mass independent indeed.}. The quantum parameter $l$ is not at the Planck scale, but goes as $m^{1/3}$, defining a ``Planck star"~\cite{Rovelli:2014cta,Barrau:2014hda,DeLorenzo:2014pta,Rovelli:2017zoa}.

\section{The dynamics of Schwarzschild-(anti)de Sitter solution}
\label{sec3}
In this section, we extend the above analysis of the interior of the Schwarzschild black-to-white hole to  Schwarzschild-(anti)de Sitter ((A)dS) spacetime with a cosmological constant,  thereby obtaining the effective internal solution of Schwarzschild-(A)dS black-to-white holes in the framework of LQG with new quantum parameters. Previously, the dynamics of Schwarzschild-(A)dS black holes with old quantum parameters have been investigated in Refs.~\cite{Brannlund:2008iw,Dadhich:2015ora,Lin:2024flv}.

\subsection{Classical dynamics of the interior Schwarzschild-(A)dS solution}
\label{sec3-1}
The Einstein-Hilbert action with a cosmological constant is given by
\begin{align}
\label{eq28}
S = \frac{1}{16\pi} \int {\text{d}}^4x \sqrt{-g} \left( R - 2\Lambda \right),
\end{align}
where $\Lambda$ is the cosmological constant. Using the same algebra, one can derive the Hamiltonian constraint in terms of Ashtekar variables as \cite{Dadhich:2015ora}:
\begin{align}
\label{eq29}
  \mathcal{H}^\Lambda\! = \! - \frac{N}{{2{\gamma ^2}}}\! \frac{{\operatorname{sgn} \left( {{p_c}} \right)}}{{\sqrt {\left| {{p_c}} \right|} }} \!\left[ {2bc{p_c}\! + \!\left( {{b^2} + {\gamma ^2}} \right){p_b}} \!- \!\Lambda p_b p_c \right].
\end{align}
In parallel with the Schwarzschild case, we first consider the lapse function Eq.~(\ref{eqa1}), which was studied in subsection~\ref{sec2-1}. According to~(\ref{eq28}) and Eq.~(\ref{eq29}), the equations of motion now read as
\begin{subequations}
\label{eq31}
\begin{align}
    \dot{b} & =-\frac{1}{2b} \left(b^2+1-\Lambda p_c \right), \label{eq31-a} \\
    \dot{c} & = -2c + \Lambda \frac{p_b}{b}, \label{eq31-b} \\
    \dot{p}_b & =\frac{p_b}{2 b^2}\left(b^2 - 1 + \Lambda p_c \right), \label{eq31-c} \\
    \dot{p}_c & = 2 p_c, \label{eq31-d}
\end{align}
\end{subequations}
where the ``dot" denotes the  time derivative with respect to $T$, and  without loss of generality the Immirzi parameter $\gamma$ has been set to unity here.
It is straightforward to obtain the solutions to Eq.~(\ref{eq31}) as follows:
\begin{subequations}
\label{eq32}
\begin{align}
    b &= \sqrt{\left(1-\frac{\Lambda}{3}r_h^2 \right)e^{-T}-1+\frac{\Lambda}{3}r_h^2 e^{2T}}, \label{eq32-a} \\
    c &= -\frac{\Lambda}{6} L_o r_h \left(2e^T+e^{-2T} \right) + \frac{L_o}{2 r_h} e^{-2T}, \label{eq32-b} \\
    p_b &= -L_o r_h e^{T} \sqrt{\left(1-\frac{\Lambda}{3}r_h^2 \right)e^{-T}-1+\frac{\Lambda}{3}r_h^2 e^{2T}}, \label{eq32-c} \\
    p_c &= r_h^2 e^{2 T}. \label{eq32-d}
\end{align}
\end{subequations}
Two integration constants are fixed by requiring that the black hole horizon lies at $T=0$ and by always satisfying the Hamiltonian constraint. If one introduces the time coordinate $\tau := r_h e^{T}$ and $m:=\frac{r_h}{2} \left(1-\frac{\Lambda}{3}r_h^2 \right)$, then the metric in the interior Schwarzschild-(A)dS black hole takes the following familiar form
\begin{align}
\label{eq33}
{\text{d}}{s^2}& \!= \!\! - \!{\left( {\frac{{2m}}{\tau} \!- \!1 +\! \frac{\Lambda }{3}{\tau^2}} \!\right)^{\!\! -\! 1}}\!{\text{d}}{\tau^2}\! + \!\left( {\frac{{2m}}{\tau}\! - 1 \!+ \frac{\Lambda }{3}{\tau^2}}\! \right)\!{\text{d}}{x^2}
\nonumber \\
&+ {\tau^2}{\text{d}}{\Omega ^2}.
\end{align}
The black hole horizon lies at $T=0$ and the singularity occurs at $T = - \infty$, so the black hole interior corresponds to $-\infty < T <0$. Obviously these solutions reduce to the metric of Schwarzschild black hole in~Eq.~(\ref{eqa4}) if $\Lambda=0$.

Next, to compare the results in semi-classical theory, we switch to another lapse function.  When cosmological constant is involved, one of course can also choose Eq.~(\ref{eq6+}) as the lapse function, leading to a metric with time coordinate $\tau  \in \left( {0,\sqrt {{r_h}} } \right)$. However, the metric~(\ref{eq:Rovelli_RBH_ds2}) does not have a cosmological constant version; the accustomed AdS metric that people are familiar with is the classical AdS metric~(\ref{eq33}) with time coordinate $\tau  \in \left( {0,{r_h}} \right)$. Therefore, to compare the results with Eq.~(\ref{eq33}), we ultimately choose  a new lapse function as follows
\begin{align}
\label{eq34}
    \hat N \left({\tau_\Lambda} \right) = \frac{\gamma \operatorname{sgn} (p_c)}{b}.
\end{align}
The corresponding canonical equations of motion  are then
\begin{subequations}
\label{eq35}
\begin{align}
    \dot{b} & =-\frac{1}{2b\sqrt{p_c}} \left(b^2+1-\Lambda p_c \right), \label{eq35-a} \\
    \dot{c} & =\frac{1}{\sqrt{p_c}}\left(-2c + \Lambda \frac{p_b}{b}\right), \label{eq35-b} \\
    \dot{p}_b & =\frac{p_b}{2 b^2 \sqrt{p_c}}\left(b^2 - 1 + \Lambda p_c \right), \label{eq35-c} \\
    \dot{p}_c & = 2 \sqrt{p_c}, \label{eq35-d}
\end{align}
\end{subequations}
where the ``dot" denotes the time derivative with respect to $\tau_\Lambda$, and $H \approx 0$ is used to simplify the equations. Between these two coordinate systems, the time coordinate transforms as
\begin{align}
    \frac{{\text{d}} {\tau_\Lambda}}{{\text{d}} T} = \frac{N}{\hat N} & =\sqrt{p_c}=r_h e^{T},\\
    {\tau_\Lambda}=r_h e^{T}\quad &\text{and} \quad T = \ln \left( {{{{\tau _\Lambda }} \mathord{\left/
 {\vphantom {{{\tau _\Lambda }} {{r_h}}}} \right.
 \kern-\nulldelimiterspace} {{r_h}}}} \right).
\end{align}
The solutions are given by
\begin{subequations}
\label{eq36}
\begin{align}
    b &= \sqrt{\frac{2m}{{\tau_\Lambda}}-1+\frac{\Lambda}{3}\tau_\Lambda^2},  \label{eq36-a}\\
    c &= L_o \left(\frac{m}{\tau_\Lambda^2}-\frac{\Lambda}{3}{\tau_\Lambda} \right),  \label{eq36-b}\\
    p_b &= -L_o {\tau_\Lambda} \sqrt{\frac{2m}{\tau_\Lambda}-1+\frac{\Lambda}{3}\tau_\Lambda^2},  \label{eq36-c}\\
    p_c &= \tau_\Lambda^2. \label{eq36-d}
\end{align}
\end{subequations}
Substituting Eq.~(\ref{eq36}) into expression~(\ref{eq4}), one obtains the line element~(\ref{eq33}) once again.
Now, the singularity occurs at $\tau_\Lambda=0$, and the black hole interior corresponds to $0 < {\tau_\Lambda} < r_h$.

\subsection{Effective dynamics of the interior Schwarzschild-(A)dS black-to-white hole}
\label{sec3-2}
The effective theory for the evolution of the interior of the Schwarzschild-(A)dS black hole can be introduced similarly to the Schwarzschild case. In this subsection, the effective dynamics  will be considered based on the Hamiltonian with the modification of the lapse function~(\ref{eq34}). Classically, the interior of the  black hole corresponds to  $0 < {\tau_\Lambda} < r_h$. By replacement ${b \to {{\sin \left( {{\delta _b}b} \right)} \mathord{\left/
 {\vphantom {{\sin \left( {{\delta _b}b} \right)} {{\delta _b}}}} \right.
 \kern-\nulldelimiterspace} {{\delta _b}}},c \to {{\sin \left( {{\delta _c}c} \right)} \mathord{\left/
 {\vphantom {{\sin \left( {{\delta _c}c} \right)} {{\delta _c}}}} \right.
 \kern-\nulldelimiterspace} {{\delta _c}}}}$, the resulting effective Hamiltonian is given by
\begin{align}
\label{eq37}
&\mathcal{H}^\Lambda_{\text{eff}}(\hat N) =  - \frac{\delta_b}{{2G{\gamma} \sin(\delta_b b) \sqrt{p_c}}}
\nonumber \\
&\times\left[ 2 \db \!\dc {p_c} + \left(\! \frac{\sin^2(\delta_b b)} {\delta_b^2} + {\gamma ^2}\! \right){p_b} - \Lambda p_b p_c  \right].
\end{align}
This Hamiltonian leads to the equations of motion:
\begin{subequations}
\label{eq:eff_eom_Ads}
\begin{align}
&\dot{b} = -\frac{\delta_b}{2\sqrt{p_c}\sin \left(\delta_b b \right)}\left(\frac{\sin^2(\delta_b b)}{\delta_b^2} + \gamma^2 - \Lambda p_c\right)\label{eq:eff_eom_Ads_b}, \\
&\dot{c} =\frac{1}{\sqrt{p_c}}\left(-2\dc+\Lambda p_b\frac{\delta_b}{\sin \left(\delta_b b \right)}\right) , \\
&\dot{p_b}= \frac{p_b \cos \left(\delta_b b \right)\delta_b^2}{2 \sqrt{p_c}\sin^2 \left(\delta_b b \right)}\left(\frac{\sin^2 \left(\delta_b b \right)}{\delta_b^2}+\Lambda p_c-\gamma^2\right), \\
&\dot{p_c} = 2 \sqrt{p_c} \cos \left(\delta_c c \right) \label{eq:eff_eom_Ads_pc},
\end{align}
\end{subequations}
where the Hamiltonian constraint $\mathcal{H}^\Lambda_{\text{eff}} \approx 0$ is used to simplify the equations of motion. Note that these equations reduce to Eq.~(\ref{eq35}) in the classical limit $\delta_b$, $\delta_c \rightarrow 0$.

Due to the difficulty of solving these equations analytically, a numerical analysis of these equations is performed. Firstly, we remark that in the presence of the cosmological constant the setup of $\delta_b$ and $\delta_c$ may be relevant to the value of $\Lambda$ in general case; thus, we intend to provide a strategy to fix $\delta_b$ and $\delta_c$ for a given cosmological constant numerically. The key  point is to require that the area enclosed by two plaquettes on the transition surface
remains to be the area gap $\Delta$, which can be implemented by  an iterative method. We elaborate on this strategy as follows. According to Eq.~(\ref{eqx1}), one gets
\begin{align}
\label{eq:delta_bc_expr}
{\delta _b} = \sqrt {\frac{\Delta }{{{{\left. {4\pi {p_c}\left( {{\delta _b},{\delta _c}} \right)} \right|}_\mathcal{T}}}}} ,\quad {\delta _c} = \frac{\Delta }{{2\pi {\delta _b}{{\left| {{p_b}\left( {{\delta _b},{\delta _c}} \right)} \right|}_\mathcal{T}}}}.
\end{align}
We require that the values of $\delta_b$ and $\delta_c$ satisfy the above equations in the presence of the cosmological constant. However, it is noted that in general case the values of $p_b$ and $p_c$ on the transition surface are unknown since we have no analytic solutions for them as in Schwarzschild case. Thus, numerically, we first set initial values $\delta_{b,0}$ and $\delta_{c,0}$ for $\delta_b$ and $\delta_c$ as those without the cosmological constant, and then plug them into the equations of motion with the cosmological constant and numerically find the solutions for $p_b$ and $p_c$ on the transition surface. Subsequently, we substitute $p_b$ and $p_c$  into Eq.~(\ref{eq:delta_bc_expr}) and obtain the values of $\delta_{b,1}$ and $\delta_{c,1}$ as the new values of $\delta_b$ and $\delta_c$ for the next iteration. Concisely, we may apply the following iterative equations:
\begin{align}
&\delta_{b,n} = \sqrt{\frac{\Delta}{4\pi p_c \left(\delta_{b,n-1},\delta_{c,n-1}\right)\left|_\mathcal{T}\right.}},
\nonumber \\
&{\delta _{c,n}} = \frac{\Delta }{{2\pi {\delta _{b,n - 1}}{{\left| {{p_b}\left( {{\delta _{b,n - 1}},{\delta _{c,n - 1}}} \right)} \right|}_{{\cal T}}}}},
\end{align}
and we stop the iteration until the area enclosed by two plaquettes on the transition surface $\mathcal{T}$ is very close to the area gap $\Delta$ (numerically we require $\sigma:=\left( \left|\Delta_n-\Delta \right| \right)/\Delta < 10^{-3}$), where $p_c \left(\delta_{b,n-1},\delta_{c,n-1} \right)$ and $p_b \left(\delta_{b,n-1},\delta_{c,n-1} \right)$ are understood as the solutions to the equations of motion with $\delta_{b,n-1}$ and $\delta_{c,n-1}$.

\begin{figure}[H]
\centering
\includegraphics[width=0.5 \textwidth]{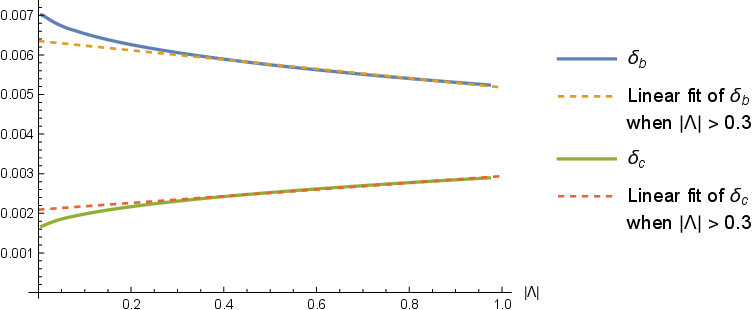}
\caption{ Relationships between $\delta_b$, $\delta_c$ and $\Lambda$ in the Schwarzschild-AdS case. $\delta_b$ and $\delta_c$ can approximately be fitted by linear functions for $\Lambda$ when $|{\Lambda}| > 0.3$. The linear fit function of $\delta_b$ is $\delta_b \approx 0.00635 - 0.00117|{\Lambda}|$ while the linear fit function for $\delta_c$ is $\delta_c \approx 0.00209 + 0.000847 |{\Lambda}|$.}
\label{figure4}
\end{figure}

Next, we apply this strategy to the Schwarzschild-AdS case with a negative cosmological constant. It is worth emphasizing that in numerical simulation and in the Planck units, $\Lambda=-1$ leads to a very large vacuum energy density that lies at the Planck energy density level. Therefore, we may restrict our analysis within the range $-1<\Lambda < 0$. In addition, in the effective region of QG, we are mainly concerned with the large mass of black holes. Without loss of generality, we still fix the mass as $m=10^6$. In Fig.~\ref{figure4}, we show the variation of $\delta_b$ and $\delta_c$ with $\Lambda$. It is observed that $\delta_b$ becomes smaller with the increase of $|{\Lambda}|$ but $\delta_c$ becomes larger. Specifically, when $|{\Lambda}| > 0.3$, both $\delta_b$ and $\delta_c$ can approximately be fitted by linear functions of $\Lambda$.
\begin{figure*}[htbp]
\centering
\includegraphics[width=0.73 \textwidth]{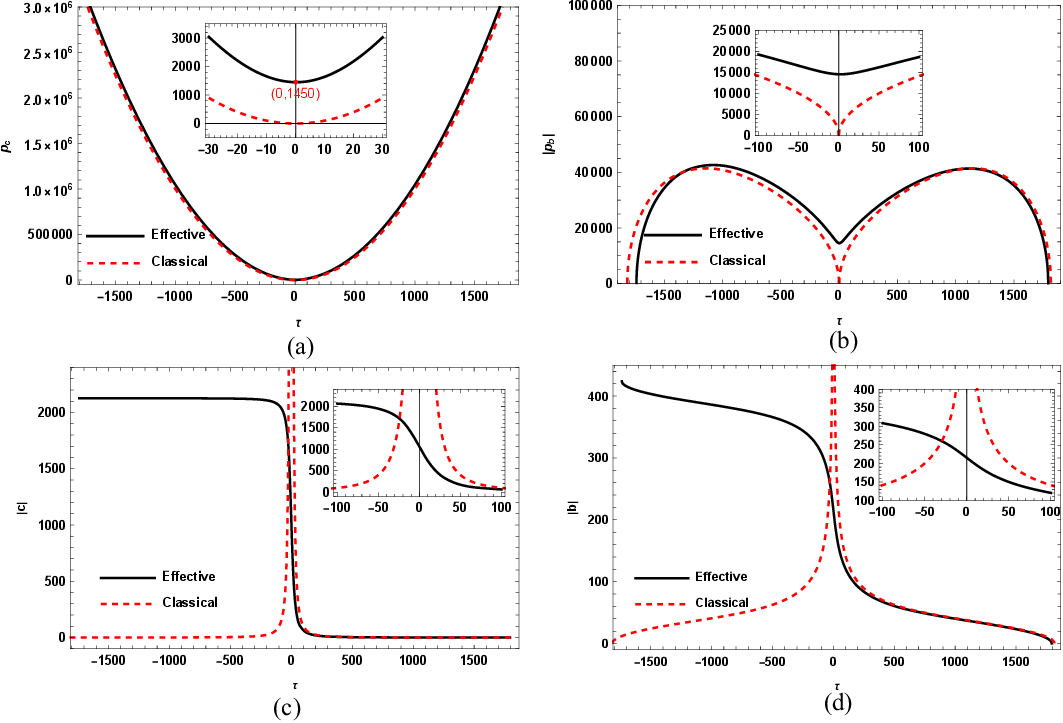}
\caption{ Time ($\tau_\Lambda$) evolution of $p_c$, $p_b$, $c$ and $b$ in the quantum corrected Schwarzschild-AdS interior for $m=10^6$, $\Lambda=-10^{-3}$, $L_o=\Delta=\gamma=1$, $\delta_b=7.410 \times 10^{-3}$ and $\delta_c = 1.476 \times 10^{-3}$. (a) $p_c$ versus $\tau$. (b) $\left|{p_b} \right|$ versus $\tau$. (c) $\left| c \right|$ versus $\tau$. (d) $\left| b \right|$ versus $\tau$. The black solid curve represents the effective Schwarzschild-AdS black-to-white hole case, while the red dashed curve denotes the classical Schwarzschild-AdS black hole case. The horizon of classical Schwarzschild-AdS black hole lies at $r_h\approx 1817.12$.}
\label{figure5}
\end{figure*}
\begin{figure*}[htbp]
\centering
\includegraphics[width=0.73 \textwidth]{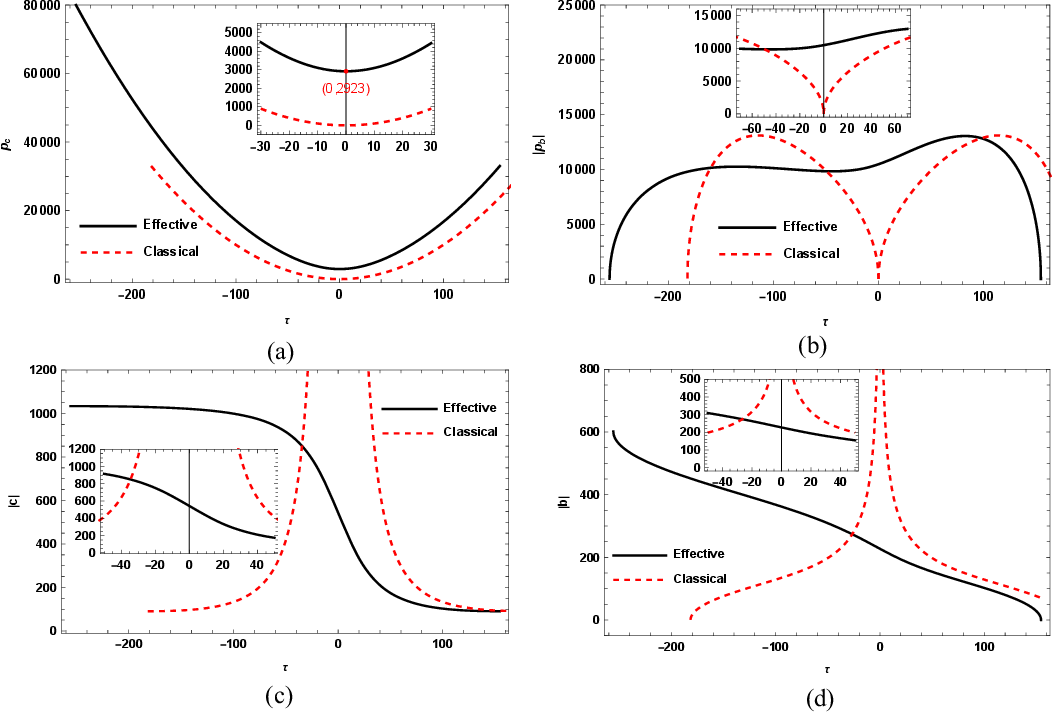}
\caption{ Time ($\tau_\Lambda$) evolution of $p_c$, $p_b$, $c$ and $b$ in the quantum corrected Schwarzschild-AdS interior for $m=10^6$, $\Lambda=-1$, $L_o=\Delta=\gamma=1$, $\delta_b=5.216\times10^{-3}$ and $\delta_c = 2.913\times10^{-3}$. (a) $p_c$ versus $\tau$. (b) $\left|{p_b} \right|$ versus $\tau$. (c) $\left| c \right|$ versus $\tau$. (d) $\left| b \right|$ versus $\tau$. The black solid curve represents the effective Schwarzschild-AdS black-to-white hole case, while the red dashed curve denotes the classical Schwarzschild-AdS black hole case. The horizon of classical Schwarzschild-AdS black hole lies at $r_h\approx181.7$.}
\label{figure6}
\end{figure*}
\begin{figure*}[htbp]
\centering
\includegraphics[width=0.73 \textwidth]{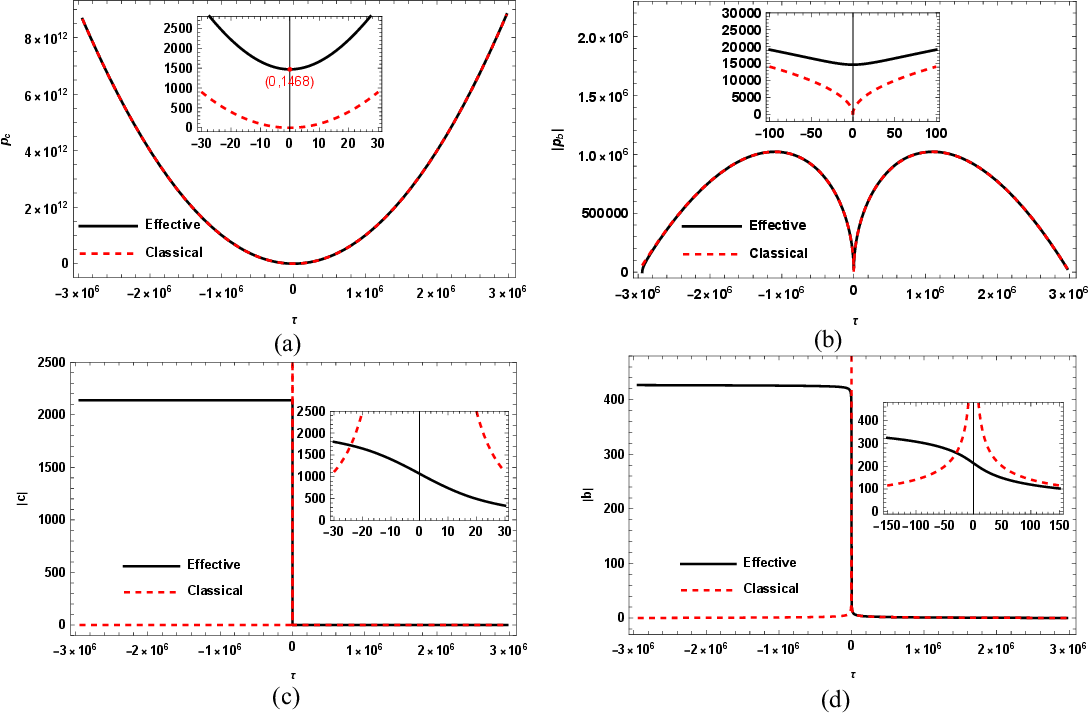}
\caption{Time ($\tau_\Lambda$) evolution of $p_c$, $p_b$, $c$ and $b$ in the quantum corrected Schwarzschild-dS interior for $m=10^6$, $\Lambda  = \left( {1/9} \right) \times {10^{ - 12}}$, $L_o=\Delta=\gamma=1$, ${\delta _b} = 7.362 \times 10^{-3}$ and $\delta_{c}= 1.468 \times 10^{-3}$. (a) $p_c$ versus $\tau$. (b) $\left|{p_b} \right|$ versus $\tau$. (c) $\left| c \right|$ versus $\tau$. (d) $\left| b \right|$ versus $\tau$. The black solid curve represents the effective Schwarzschild-dS black-to-white hole case, while the red dashed curve denotes the classical Schwarzschild-dS black hole case. The horizon of classical Schwarzschild-dS black hole lies at $r_h=3\times10^6$.}
\label{figure7}
\end{figure*}

Now, we illustrate the specific solutions for the case with large mass and small cosmological constant. Let us set $m=10^6$ and $\Lambda=-10^{-3}$. By the iteration method, we find the quantum parameters should be set as $\delta_b=7.410 \times 10^{-3}$ and $\delta_c = 1.476 \times 10^{-3}$ to guarantee that the area enclosed by two plaquettes on the transition surface $\mathcal{T}$ is the area gap $\Delta$. In comparison with the values in the Schwarzschild case which are $\delta_b=7.362 \times 10^{-3}$ and $\delta_c = 1.468 \times 10^{-3}$, we notice that both quantities just have a small shift. Correspondingly,  the evolution of $p_c$, $p_b$, $c$, and $b$ are obtained by solving the equations of motion, and the results are illustrated in Fig.~\ref{figure5}. It is noticed that the evolution of all the variables can pass through $\tau_\Lambda = 0$, signaling that the original singularity at  $\tau_\Lambda = 0$ disappears. The variable $p_c$ in Fig.~\ref{figure5}(a) takes the minimal value $p_c\left|_{\text{min}}\right.\approx 1450$ at  $\tau_\Lambda = 0$,
which is slightly lower than the value in the Schwarzschild case $ \left(p_c\left|_{\text{min}}\right. \right)_{\text{Sch}}=\gamma L_o m \delta_c \approx 1468$, but still of the same order of magnitude. Therefore, in this case the presence of the negative cosmological constant does not significantly alter the minimal value of $p_c$, and thus the radius of the transition surface $l=\sqrt{p_c\left|_{\text{min}}\right.}$ also scales as $m^{1/3}$. On the contrary, we point out that the radius of the classical Schwarzschild-AdS black hole $r_h=\left(-6m/\Lambda \right)^{1/3}+\mathcal{O}\left(m^{-1/3} \right)\approx 1817.12$ also scales as $m^{1/3}$ and thus is dramatically less than the radius of the classical Schwarzschild black hole with the same mass since $ \left(r_h \right)_{\text{Sch}}=2m=2\times10^6$. It is noted that $l$ and $r_h$ are of the same order of magnitude as the mass, implying that the transition surface is closer to the horizon in the large $m$ limit in comparison with that in the Schwarzschild case, and the effective dynamics is significantly contrasts with the classical dynamics in the interior of the Schwarzschild-AdS black hole. In addition, according to Fig.~\ref{figure5}(b), it can be found that the classical case (red dashed curve) is strictly symmetric with respect to $\tau = 0$, whereas the quantization case (black solid curve) is not, which suggests that the effective theory with a cosmological constant after quantization breaks the symmetry of spacetime.

Then, we consider the specific solutions with a large mass and a large cosmological constant with $\Lambda=-1$ and $m=10^6$. In this case, we find $\delta_b=5.216\times10^{-3}$ and $\delta_c = 2.913\times10^{-3}$ using the the iteration method, and the evolution of spacetime is illustrated in Fig.~\ref{figure6}. It is worth noting that the minimal value of $p_c$ is larger than the value in the $\Lambda=-10^{-3}$ case, which means the transition surface is closer to the event horizon  in the extreme case. In fact, it is found that $l=\sqrt{p_c\left|_{\text{min}}\right.}\approx 54.0$, while $r_{\text{BH}} = \sqrt{p_c\left|_{\text{BH}}\right.} \approx179.8$, so that we have $l \approx 0.3
 r_{\text{BH}}$. Furthermore, the radius of the white hole horizon $r_{\text{WH}} = \sqrt{p_ c\left|_{\text{WH}}\right.} \approx 286.0$ is significantly larger than $r_{\text{BH}}$. We recall that, in the Schwarzschild case with the same mass $m=10^6$, the black hole mass agrees with the white hole mass, namely $m_{\text{BH}} = m_{\text{WH}} + \mathcal{O} \left(1\right)$. This relation does not hold when $\Lambda = -1$, and the negative cosmological constant disrupts the approximate symmetry between the black hole and white hole. This can also be seen from Fig.~\ref{figure6}(a) and Fig.~\ref{figure6}(b): these two figures exhibit  manifest asymmetry between the black hole region and the white hole region. Of course, we remark that, in general, this discrepancy depends on the ratio of the mass of black hole to the square root of the cosmological constant, namely $m/\sqrt{|\Lambda|}$. Numerically, we find that when $m/\sqrt{|\Lambda|} >10^8$, the effect of the cosmological constant is limited and the dynamics is similar to the Schwarzschild case, while if $m/\sqrt{|\Lambda|} <10^6$, then the effect of the cosmological constant becomes notable, and may leads to manifest asymmetry between the black hole region and the white hole region.

Next, we turn to consider the Schwarzschild-dS black holes with a positive cosmological constant, which is slightly different from the Schwarzschild-AdS case. The classical Schwarzschild-dS spacetime has two horizons when $0<9\Lambda m^2 <1$, one degenerate horizon when $9 \Lambda m^2 = 1$, and no horizon when $9 \Lambda m^2 > 1$. For simplicity, we consider $9 \Lambda m^2 = 1$, and similar conclusions can be obtained for the case where $0<9\Lambda m^2 <1$. When $9 \Lambda m^2 = 1$, the event horizon lies at ${r_h} = 3m = 1/\sqrt \Lambda$. This is the so-called Nariai black hole~\cite{Nariai1951OnAN}. This black hole exhibits distinct evolutionary behavior with different values of mass. One extremal case is when the mass of the black hole is very large; for instance, by setting the mass of the black hole to be $10^6$, then $\Lambda  = \left( {{1 \mathord{\left/ {\vphantom {1 9}} \right. \kern-\nulldelimiterspace} 9}} \right) \times {10^{- 12}}$. The evolution of such  a black-to-white hole is illustrated in Fig.~\ref{figure7}. Similarly, by requiring that the quantum parameters  ${\delta _b}$ and ${\delta_c}$ satisfy relation~(\ref{eqx1}) and employing the iterative method, we find that  ${\delta _b} = 7.362 \times 10^{-3}$ and $\delta_{c}= 1.468 \times 10^{-3}$, which are almost the same values as those given by Eq.~(\ref{eq:Ashtekar_delta_bc}) for $m=10^6$. This arises from the fact that the cosmological constant is extremely small, such that its effects may be neglected. Therefore, in the large mass limit the effective dynamics of the Schwarzschild-dS black-to-white hole is very similar to the effective dynamics of the Schwarzschild black-to-white hole.

\subsection{The symmetry of the geometry between black hole and white hole}
\label{sec3-3}
In the previous subsection, it was demonstrated that the presence of a cosmological constant may lead to an asymmetry between the black hole region and the white hole region and its impact depends on the ratio $m/\sqrt{\Lambda}$. In section~\ref{sec2-3}, it is shown that in the Schwarzschild case, if $\delta_b$ and $\delta_c$ satisfy the condition $\cos \left(\delta_b b \right)=0$ at the transition surface $\mathcal{T}$, the metric is exactly symmetric between the black hole region and the white hole region. In this subsection, we intend to prove that this conclusion holds true even in the presence of a cosmological constant. In fact, one can show that $p_b\left(\tau_\Lambda \right)$ and $p_c\left(\tau_\Lambda \right)$ are even functions, whereas $b\left(\tau_\Lambda\right)$ and $c\left(\tau_\Lambda\right)$ are odd functions if $\cos \left(\delta_b b \left(0 \right)\right)=0$ is satisfied. Thus, the metric, which depends on $p_b$ and $p_c$, is also an even function, and remains invariant under the transformation $\tau \rightarrow -\tau$.

The proof can be demonstrated as follows. In order to show that $b$ and $c$ are odd functions, while $p_b$ and $p_c$ are even functions, we need to prove the following equations:
\begin{subequations}
\begin{gather}
b \left({\tau_\Lambda} \right) + b \left(-{\tau_\Lambda} \right)=2 b \left(0 \right),\\
c \left({\tau_\Lambda} \right) + c \left(-{\tau_\Lambda} \right)=2 c \left(0 \right),\\
p_b \left({\tau_\Lambda} \right) = p_b \left(-{\tau_\Lambda} \right),\\
p_c \left({\tau_\Lambda} \right) = p_c \left(-{\tau_\Lambda} \right).
\end{gather}
\end{subequations}
Then, by defining $\tilde{b} \left({\tau_\Lambda} \right) = 2b \left(0 \right) - b \left(-{\tau_\Lambda} \right)$, $\tilde{c}\left(\tau \right) = 2c \left(0 \right) - c \left(-{\tau_\Lambda} \right)$, $\tilde{p}_b \left({\tau_\Lambda} \right)=p_b \left(-{\tau_\Lambda} \right)$ and $\tilde{p}_c \left ({\tau_\Lambda} \right)=p_c \left(-{\tau_\Lambda} \right)$, we need to prove that $\tilde{b} \left({\tau_\Lambda} \right)=b \left({\tau_\Lambda} \right)$, $\tilde{c} \left({\tau_\Lambda} \right)=c \left({\tau_\Lambda} \right)$, $\tilde{p}_b \left({\tau_\Lambda})=p_b({\tau_\Lambda} \right)$ and $\tilde{p}_c \left({\tau_\Lambda} \right)=p_c \left ({\tau_\Lambda} \right)$. This can be shown by demonstrating that they are the same set of solutions to Eq.~(\ref{eq:eff_eom_Ads}). Actually, based on the condition that $\cos \left(\delta_b b \right)=0$ at $\mathcal{T}$, one can derive the following corollary
\begin{align}
    \cos \left(\delta_b b \left(0 \right)\right) = 0 \quad \text{and} \quad \delta_b b \left(0\right) = n_b \pi + \frac{\pi}{2}.
\end{align}
From Eq.~(\ref{eq:eff_eom_Ads_pc}), since $p_c$ takes the minimum value at $\mathcal{T}$, one obtains
\begin{align}
    \cos \left(\delta_c c \left(0\right)\right) = 0 \quad \text{and} \quad \delta_c c \left(0\right) = n_c \pi + \frac{\pi}{2}.
\end{align}
Thus, substituting  $\tilde{b} \left({\tau_\Lambda} \right)$, $\tilde{c}\left({\tau_\Lambda} \right)$, $\tilde{p}_b \left({\tau_\Lambda} \right)$ and $\tilde{p}_c \left({\tau_\Lambda} \right)$  into the right-hand side of Eq.~(\ref{eq:eff_eom_Ads_b}), one gets
\begin{align}
&- \frac{{{\delta _b}}}{{2\sqrt {{{\tilde p}_c}\left( {{\tau _\Lambda }} \right)} \sin ({\delta _b}\tilde b\left( {{\tau _\Lambda }} \right))}}\left[ {\frac{{{{\sin }^2}\left( {{\delta _b}\tilde b\left( {{\tau _\Lambda }} \right)} \right)}}{{\delta _b^2}} + {\gamma ^2} - \Lambda {{\tilde p}_c}\left( {{\tau _\Lambda }} \right)} \right]
\nonumber\\
&=   -   \frac{{{\delta _b}}}{{2\sqrt {{p_c}\left( { - {\tau _\Lambda }} \right)} \sin \left[ {\pi  - {\delta _b}b\left( { - {\tau _\Lambda }} \right)} \right]}}
\nonumber\\
& \times \left\{ {\frac{{{{\sin }^2}\left[ {\pi  - {\delta _b}b\left( { - {\tau _\Lambda }} \right)} \right]}}{{\delta _b^2}} + {\gamma ^2} - \Lambda {p_c}\left( { - {\tau _\Lambda }} \right)} \right\}
\nonumber \\
&=  - \frac{{{\delta _b}}}{{2\sqrt {{p_c}( - {\tau _\Lambda })} \sin \left[ {{\delta _b}b\left( { - {\tau _\Lambda }} \right)} \right]}}
\nonumber \\
& \times  \left\{ {\frac{{{{\sin }^2}\left[ {{\delta _b}b\left( { - {\tau _\Lambda }} \right)} \right]}}{{\delta _b^2}} + {\gamma ^2} - \Lambda {p_c}\left( { - {\tau _\Lambda }} \right)} \right\} = {\left. {\frac{{{\text{d}}b}}{{{\text{d}}{\tau _\Lambda }}}} \right|_{ - {\tau _\Lambda }}}
\nonumber \\
&= {\left. {\frac{{{\text{d}}\tilde b}}{{{\text{d}}{\tau _\Lambda }}}} \right|_{{\tau _\Lambda }}},
\end{align}
which means that $\tilde{b} \left({\tau_\Lambda} \right)$ also satisfies equation Eq.~(\ref{eq:eff_eom_Ads_b}). The other three equations can be verified in  a similar manner. We have
\begin{subequations}
\begin{align}
&\frac{1}{{\sqrt {{{\tilde p}_c}({\tau_\Lambda} )} }}\left[ { - 2\frac{{\sin \left( {{\delta _c}\tilde c\left( {\tau_\Lambda}  \right)} \right)}}{{{\delta _c}}} + \Lambda {{\tilde p}_b}({\tau_\Lambda} )\frac{{{\delta _b}}}{{\sin \left( {{\delta _b}\tilde b\left( {\tau_\Lambda}  \right)} \right)}}} \right]
\nonumber \\
& = {\left. {\frac{{{\text{d}}\tilde c}}{{{\text{d}}{\tau_\Lambda} }}} \right|_{\tau_\Lambda} },\\
&\frac{{{{\tilde p}_b}({\tau_\Lambda} )\cos \left( {{\delta _b}\tilde b\left( {\tau_\Lambda}  \right)} \right)\delta _b^2}}{{2\sqrt {{{\tilde p}_c}\left({\tau_\Lambda} \right)} {{\sin }^2}\left( {{\delta _b}\tilde b\left( {\tau_\Lambda}  \right)} \right)}}\left[ {\frac{{{{\sin }^2}\left( {{\delta _b}\tilde b\left( {\tau_\Lambda}  \right)} \right)}}{{\delta _b^2}} + \Lambda {{\tilde p}_c} \left({\tau_\Lambda} \right) - {\gamma ^2}} \right]
\nonumber \\
& = {\left. {\frac{{{\text{d}}{{\tilde p}_b}}}{{{\text{d}}{\tau_\Lambda} }}} \right|_{\tau_\Lambda} }, \\
&2\sqrt {{{\tilde p}_c} \left({\tau_\Lambda} \right)} \cos \left( {{\delta _c}\tilde c\left( {\tau_\Lambda}  \right)} \right) = {\left.{\frac{{{\text{d}}{{\tilde p}_c}}}{{{\text{d}}{\tau_\Lambda}}}} \right|_{\tau_\Lambda} }.
\end{align}
\end{subequations}

The above expressions indicate that $\tilde{b} \left({\tau_\Lambda} \right)$, $\tilde{c}\left({\tau_\Lambda} \right)$, $\tilde{p}_b \left({\tau_\Lambda} \right)$, and $\tilde{p}_c \left({\tau_\Lambda} \right)$ are also solutions to Eq.~(\ref{eq:eff_eom_Ads}). Furthermore, these two sets of solutions share the same initial conditions $\tilde{b}\left(0 \right) = b \left(0 \right)$, $\tilde{c}\left(0 \right) = c \left(0\right)$, $\tilde{p}_b \left(0 \right) = p_b \left(0 \right)$, $\tilde{p}_c \left(0 \right) = p_c \left(0 \right)$, which implies that they must be identical to each other. Therefore, one obtains $\tilde{b} \left({\tau_\Lambda})=b({\tau_\Lambda} \right)$, $\tilde{c} \left({\tau_\Lambda} \right)=c \left({\tau_\Lambda} \right)$, $\tilde{p}_b \left({\tau_\Lambda})=p_b({\tau_\Lambda} \right)$, $\tilde{p}_c \left({\tau_\Lambda} \right)=p_c \left ({\tau_\Lambda} \right)$.
\begin{table}[htbp]
\footnotesize
\caption{Values of $\delta_b$ and $\delta_c$ for different Schwarzschild and Schwarzschild-(A)dS black-to-white holes. We set $m=10^6$, and $\delta_b$, $\delta_{c1}$ are obtained through iteration while $\delta_{c2}$ is obtained by fixing $\delta_b$ and then numerically solving Eq.~(\ref{eq:eff_eom_Ads}) with condition $\cos \left(\delta_b b \left(0 \right) \right)=0$.}
\label{tab:deltabc}
\doublerulesep 0.3pt \tabcolsep 1pt
\begin{tabular}{cccc}
\hline
The cosmological constant$$&$\delta_b $& $\delta_{c1}$  &$\delta_{c2}$\\
\hline
$\Lambda=0$ &$7.362 \times {10^{ - 3}} $ & $1.468\times10^{-3}$&  $1.469\times10^{-3}$\\
$\Lambda=-10^{-3}$ & $7.410\times10^{-3}$ & $1.476\times10^{-3}$ &$1.623\times10^{-3}$\\
$\Lambda = -1 $ & $5.216\times 10^{-3} $ & $2.913\times10^{-3} $ &$5.532\times 10^{-4}$ \\
$\Lambda=\left(1/9\right)\times10^{-12} $& $7.362\times10^{-3} $& $1.468\times10^{-3} $&$1.469\times10^{-3}$\\
\hline
\end{tabular}
\end{table}

So far, we have shown that $p_b(\tau_\Lambda)$ and $p_c(\tau_\Lambda)$ are even functions, whereas $b(\tau_\Lambda)$ and $c(\tau_\Lambda)$ are odd functions if $\cos \left(\delta_b b \left(0 \right)\right)=0$ is satisfied. Although we have proven the above property under a specific lapse function~(\ref{eq34}), it is worthwhile to emphasize that the proof holds true under any lapse function. As a matter of fact, the conditions $\cos \left(\delta_b b \right)=0$ at $\mathcal{T}$ do not depend on the choice of the lapse function. Using the same method, one can also prove the conclusion mentioned in section~\ref{sec2-3} that $p_b(T)$ and $p_c(T)$, which are the solutions to Eq.~(\ref{eq:sol_of_eff_eom_N1}), are even functions with respect to $T=T_\mathcal{T}$, whereas $b(T)$ and $c(T)$ are odd functions with respect to $T=T_\mathcal{T}$ if $\cos \left(\delta_b b \left(T_\mathcal{T}\right)\right)=0$.

\begin{figure*}[htbp]
\centering
\includegraphics[width=0.73 \textwidth]{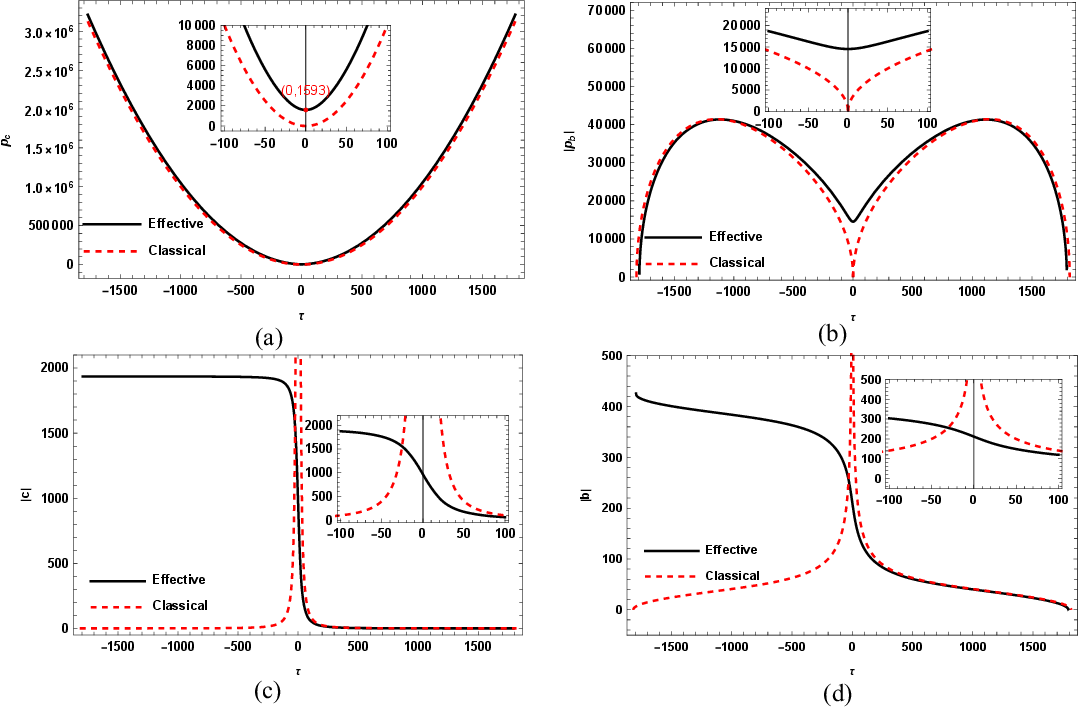}
\caption{Time ($\tau_\Lambda$) evolution of $p_c$, $p_b$, $c$ and $b$ in the quantum corrected Schwarzschild-AdS interior for $m=10^{6}$, $\Lambda=-10^{-3}$, $L_o=\Delta=\gamma=1$, $\delta_{b}\approx7.410\times10^{-3}$ and $\delta_{c2}\approx1.623\times10^{-3}$. (a) $p_c$ versus $\tau$. (b) $\left|{p_b} \right|$ versus $\tau$. (c) $\left| c \right|$ versus $\tau$. (d) $\left| b \right|$ versus $\tau$.}
\label{figure8}
\end{figure*}

\begin{figure*}[htbp]
\centering
\includegraphics[width=0.73 \textwidth]{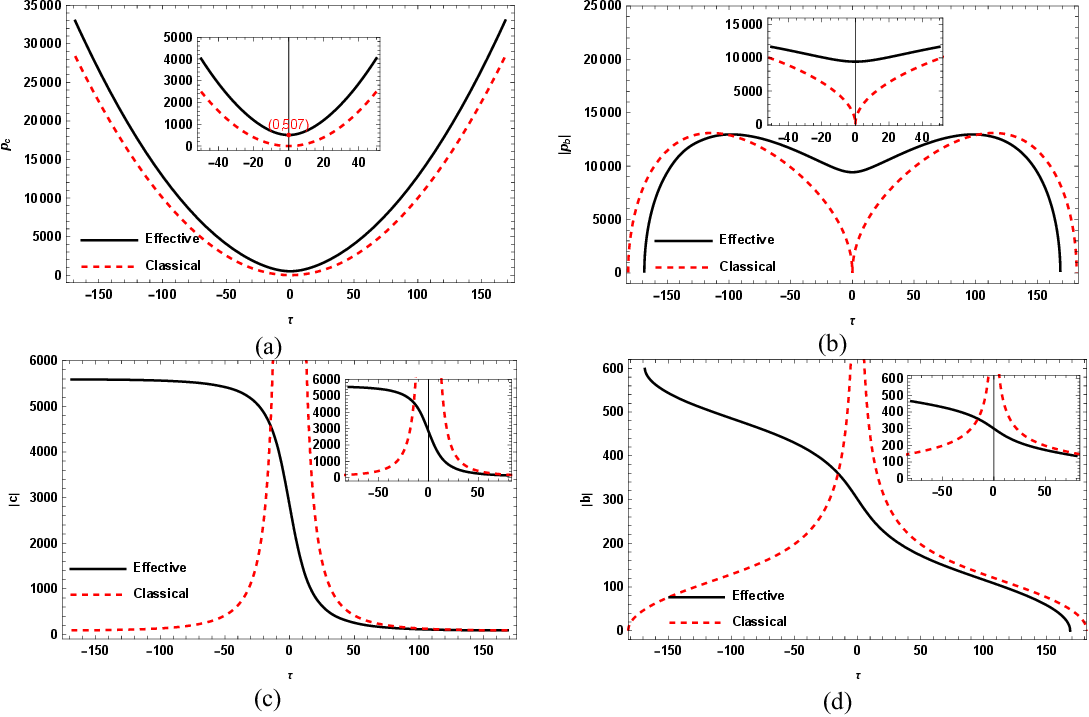}
\caption{Time ($\tau_\Lambda$) evolution of $p_c$, $p_b$, $c$ and $b$ in the quantum corrected Schwarzschild-AdS interior for $m=10^{6}$, $\Lambda=-1$, $L_o=\Delta=\gamma=1$, $\delta_{b}\approx5.216\times10^{-3}$ and $\delta_{c2}\approx5.532 \times10^{-4}$. (a) $p_c$ versus $\tau$. (b) $\left|{p_b} \right|$ versus $\tau$. (c) $\left| c \right|$ versus $\tau$. (d) $\left| b \right|$ versus $\tau$.}
\label{figure9}
\end{figure*}

\begin{figure*}[htbp]
\centering
\includegraphics[width=0.73 \textwidth]{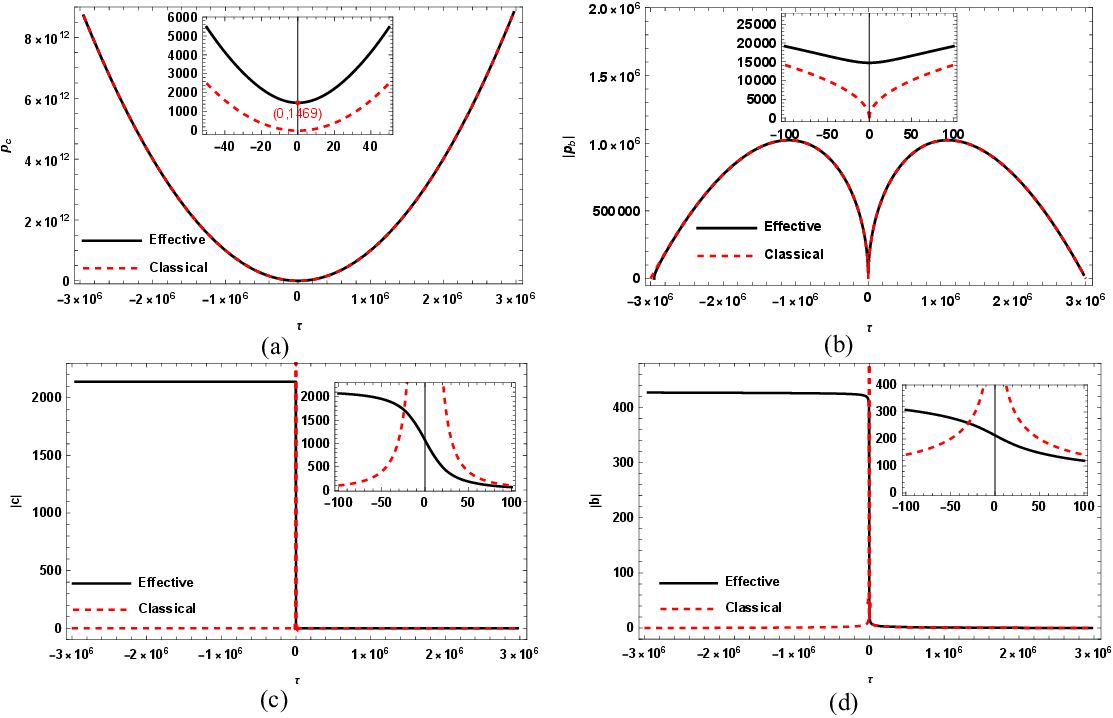}
\caption{Time ($\tau_\Lambda$) evolution of $p_c$, $p_b$, $c$ and $b$ in the quantum corrected Schwarzschild-dS interior for $m=10^6$, $\Lambda=\left({1/9}\right)\times10^{-12}$, $L_o=\Delta=\gamma=1$, $\delta_{b}\approx7.362\times10^{-3}$ and $\delta_{c2}\approx1.468\times10^{-3}$. (a) $p_c$ versus $\tau$. (b) $\left|{p_b} \right|$ versus $\tau$. (c) $\left| c \right|$ versus $\tau$. (d) $\left| b \right|$ versus $\tau$.}
\label{figure10}
\end{figure*}

Next, we discuss the value of quantum parameters that may give rise to symmetric solutions. Once $\delta_b$ is fixed, the corresponding value of $\delta_c$ leading to symmetric solutions can be obtained by numerically solving  Eq.~(\ref{eq:eff_eom_Ads}) with the condition $\cos(\delta_b b(0))=0$, which is denoted as $\delta_{c2}$. The various values of $\delta_b$ and $\delta_{c2}$ leading to symmetric solutions for different black-to-white holes are listed in Tab.~\ref{tab:deltabc}, where for comparison the value of $\delta_{c1}$ which is obtained through iteration  is also given. The time evolution of $p_c$, $p_b$, $c$, and $b$ with quantum parameters $\delta_b$ and $\delta_{c2}$ is illustrated in Figs.~\ref{figure8}-\ref{figure10}. In those figures, the black solid curve represents the effective Schwarzschild-(A)dS black-to-white hole case, while the red dashed curve denotes the classical Schwarzschild-(A)dS black hole case. It is manifestly observed that all the functions exhibit symmetric behavior as they perform in the analytic analysis.
\begin{figure*}[htbp]
\centering
\includegraphics[width=0.72 \textwidth]{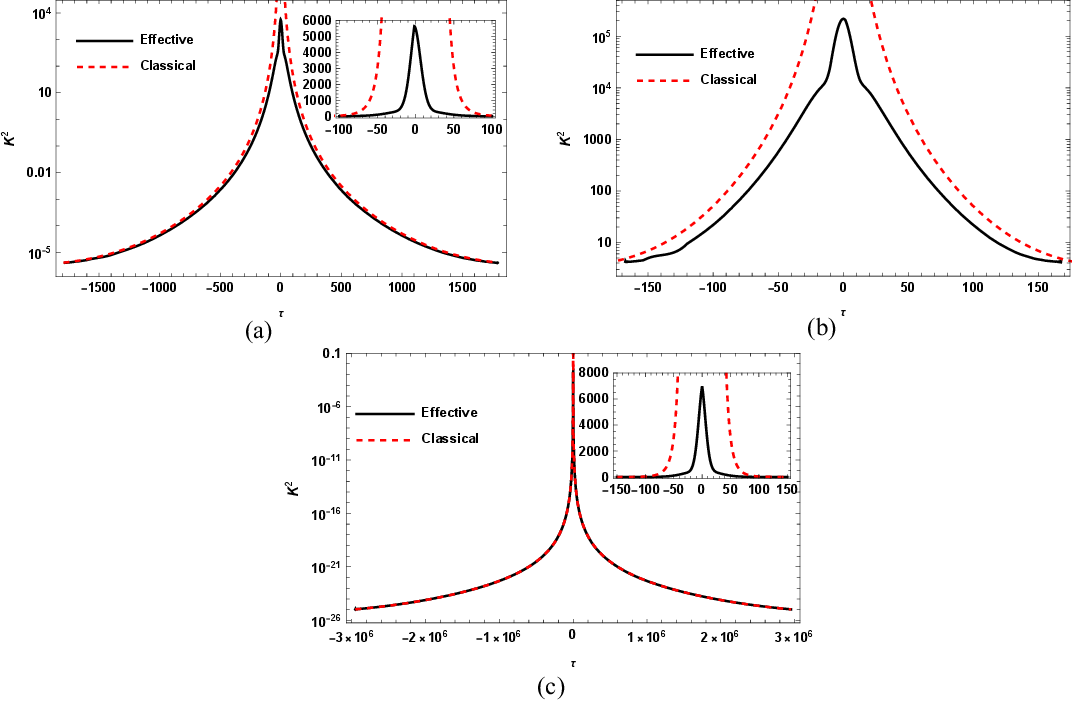}
\caption{The Kretschmann scalar in the quantum corrected Schwarzschild-(A)dS interior for (a) $m=10^6$, $\Lambda=-10^{-3}$, $\delta_{b}\approx7.410\times10^{-3}$, and $\delta_{c2} \approx1.623\times10^{-3}$. (b) $m=10^6$, $\Lambda=-1$, $\delta_{b}\approx 5.216 \times10^{-3}$, and $\delta_{c2}\approx 5.532\times 10^{-4}$. (c) $m=10^6$, $\Lambda=\left({1/9}\right)\times10^{-12}$, $\delta_{b}\approx7.362\times10^{-3}$, and $\delta_{c2}\approx1.469\times10^{-3}$. The black solid curve represents the effective Schwarzschild-(A)dS black-to-white hole case, while the red dashed curve denotes the classical Schwarzschild-(A)dS black hole case.}
\label{figure11}
\end{figure*}

About the symmetric solutions, we present the following remarks. Firstly, as the cosmological constant is relatively small compared with the mass of the black hole (numerically, the ratio $m/\sqrt{|\Lambda|} > 10^8$), we find $\delta_{c1}$ and $\delta_{c2}$ are quite close. This indicates that the symmetric solutions also satisfy the requirement that the area enclosed by two plaquettes on the transition surface $\mathcal{T}$ is the area gap $\Delta$ very well. On the other hand, as the cosmological constant is relatively large comparing with the mass of the black hole (numerically, the ratio $m/\sqrt{|\Lambda|} < 10^6$), then we find $\delta_{c1}$ and $\delta_{c2}$ are not close (see the data for $\Lambda=-1$ in Tab.~\ref{tab:deltabc}). In this case, we find that the area enclosed by two plaquettes on the transition surface $\mathcal{T}$ has a discrepancy with the area gap $\Delta$ (for $\Lambda=-1$, $\Delta_{\text{symmetry}}\approx 0.17\Delta$). Secondly, we have adopted the method of finding $\delta_{c2}$ by fixing $\delta_{b}$ to search for symmetric solutions. Alternatively, one may adjust both parameters $\delta_{b}$ and $\delta_{c}$ to comply with the symmetric condition. Then for a large cosmological constant, the discrepancy between the area of the plaquette on the transition surface and the area gap $\Delta$ would become smaller. Thirdly, it is interesting to compare our results with those in recent Refs.~\cite{Alonso_Bardaji_2023,belfaqih2024blackholeseffectiveloop,mato2024sphericallysymmetricloopquantum}, where spherically symmetric solutions with a cosmological constant have also been constructed but  in a covariant manner. It is found that the metrics in these references have different spacetime structures compared to ours. We point out that this difference arises from the different choices of the $\mu$-scheme. Nevertheless, we remark that the symmetric solutions in our paper share the same spacetime structure and similar properties with the metric obtained by using the covariant $\bar{\mu}$-scheme (C.~Zhang, et. al., by private communication.), where a cosmological constant is taken into account and the radius of the transition surface of the covariant metric also scales as $m^{-\frac{1}{3}}$, implying that the transition surface is closer to the event horizon in the presence of a cosmological constant.

Finally, we discuss the Kretschmann scalars of the quantum corrected Schwarzschild-(A)dS black-to-white holes, which  are plotted  for various quantum parameters in Fig.~\ref{figure11}. It can be seen that the Kretschmann scalars are even functions due to the symmetry of the metric. They are finite in the entire interior region and reach a single maximum at $\mathcal{T}$, indicating that the classical singularity is removed and matter is allowed to travel from the black hole to the white hole.

\section{Conclusions and discussions}
\label{sec4}
In this paper, we have investigated the dynamics of the black-to-white holes in the interior of the event horizon in the framework of LQG, where an effective theory with holonomy corrections is proposed at the the semi-classical level. This work has made progress on the following topics.  Firstly, we have explicitly shown that for any system with a Hamiltonian constraint, different choices of the lapse function lead to the same dynamics, even at the semi-classical level, where the canonical equations of motion are modified due to the quantum correction. Of course, this result is not surprising since we know that the Hamiltonian constraint should lead to a re-parameterization invariance for the time coordinate. Secondly, inspired by this observation, we have looked into the dynamics of the Schwarzschild black-to-white hole with different choices of the lapse function and explicitly demonstrated that the black-to-white solution with the metric in~(\ref{eq:Rovelli_RBH_ds2}), which is proposed based on the phenomenological consideration, can be viewed as the coarse-grained version of the black-to-white hole solution with the metric in~(\ref{eq:Sch_eff_ds2}), since both metrics exhibit the same behavior at the leading-order expansion near the transition surface. In this sense, we have  provided the theoretical foundation for the metric in Eq.~(\ref{eq:Rovelli_RBH_ds2}) by embedding it into the framework of LQG and treating it as the approximation or coarse-grained version of the  solution to the equations of motion in the semi-classical theory. Thirdly, inspired by the exact symmetry appearing in  the metric described by Eq.~(\ref{eq:Rovelli_RBH_ds2}), we have constructed the black-to-white hole solutions with exact symmetry between the black hole region and the white hole region by properly fixing the quantum parameters in the effective theory of LQG. Thus, the geometry of the black hole and white hole is symmetric, needless to worry about the amplification of mass that might arise due to the mass difference between the black hole and the white hole. Remarkably, we have also demonstrated that with an appropriate  choice of lapse function (or the time coordinate),  the transition surface from the black hole to the white hole is exactly located at $\tau = 0$, which previously indicated the presence of the singularity, as demonstrated in Fig.~\ref{figure2}(b). In this sense, the picture of replacing the singularity with a bounce from the black hole to the white hole is more intuitive and vivid in comparison with the previous work where different lapse functions are employed. As a result, in this coordinate system, the dynamics of the interior black-to-white hole is plotted with a picture analogous to that in loop quantum cosmology, where a cosmological singularity is replaced by the big bounce at the same time moment.  The analogy between the dynamics of the black hole interior and that of the universe has been revealed more evidently.

Subsequently, the effective theory has been applied to investigate the interior dynamics of a black hole in the presence of a cosmological constant. We have proposed an iteration method to fix the values of quantum parameters to guarantee that the area enclosed by two plaquettes on the transition surface remains the area gap. By numerically solving the equations of motion, we demonstrate that quantum gravity effects can eliminate the singularity in spacetime and extend time-like coordinates from the black hole region to the white hole region, thus providing the possibility for particles to reach other universes. More importantly, it is found that the geometry, as well as the  physical properties of the black hole interior, is different from that of the Schwarzschild interior. For the Schwarzschild-AdS black holes, the radius of the event horizon of the black hole  is not the same as that of the white hole because  the presence of a negative cosmological constant  breaks this symmetry. More importantly,  the locations of both the transition surface and the event horizon scale as $m^{1/3}$ for larger $m$, indicating that spacetime bounces near the event horizon. On the other hand, for Schwarzschild-dS black-to-white holes, the time evolution of of $p_c$, $p_b$, $c$, and $b$ is similar to those of quantities in the Schwarzschild black-to-white case (see Fig.~\ref{figure7}),  suggesting that both of them have similar dynamical properties. Finally, in order to fix the asymmetric problem of the Schwarzschild (A)dS metric, a scheme has been proposed to modify the quantum parameters $\delta_c$ and $\delta_b$. It is shown that when the quantum parameters satisfy the relation $\cos \left(\delta_b b \left(\tau_\mathcal{T} \right)\right)=0$, the symmetry between the black hole region and the white hole region  may be restored. In addition, as the ratio of $m/\sqrt{\Lambda} > 10^8$, the symmetric solutions are subject to the condition that the area enclosed by two plaquettes on the transition surface remains to be the area gap.

An open problem in the minisuperspace quantization method is that it is not clear whether there exists a covariant action whose symmetry reduction yields these dynamical equations. This issue  was initially investigated in Ref.~\cite{Bojowald_2012}. Subsequently, some models that adhere to the general covariance were proposed in Refs.~\cite{zhang2024blackholescovarianceeffective,Alonso_Bardaji_2022,Alonso_Bardaji_2022_2,Bojowald_2024_1,Bojowald_2024_2}. The embedding of our model into a covariant formalism is a meaningful question, and several possible approaches for constructing a covariant action are discussed  in Ref.~\cite{Ashtekar_2020}.
In this paper we have demonstrated that the symmetric metric, which is obtained using the new $\mu$-scheme, is consistent with the covariant spherically symmetric metric under the coarse-graining approximation. Thus, the construction of symmetric solutions may provide an efficient way to link the models with minisuperspace quantization to those in the covariant approach.

In the end, we remark that we have only focused on the interior dynamics of the black hole in this paper. The effective spacetime structure outside the event horizon and how it matches the interior region should be studied in detail, and we leave this for further investigation.

\vspace*{3.0ex}
{\bf Acknowledgements}
\vspace*{1.0ex}

We are very grateful to Guoping Li, Kai Li, Pan Li, Yongge Ma, Jun Nian, Wenbin Pan, Yu Tian and Hongbao Zhang for helpful discussions. Specially we thank Xiangdong Zhang and Cong Zhang for their correspondence with detailed comments and suggestions, which help us to improve the manuscript greatly. This work is supported in part by the Natural Science Foundation of China (Grant Nos.~12035016,~12275275,~12105231 and~12275350). It is also supported by Beijing Natural Science Foundation (Grant No.~1222031) and the Innovative Projects of Science and Technology ( E2545BU210) at IHEP, and by Natural Science Foundation of Sichuan Province (Grant No. 2024ZYD0075).


\begin{thebibliography}{200}
\providecommand{\url}[1]{\texttt{#1}}
\providecommand{\urlprefix}{URL }
\providecommand{\doi}[1]{doi:~\href{http://doi.org/#1}{\nolinkurl{#1}}}
\providecommand{\arXiv}[1]{\href{https://arxiv.org/abs/#1}{\nolinkurl{https://%
arxiv.org/abs/#1}}}
\providecommand{\eprint}[1]{\href{http://arxiv.org/abs/#1}{\nolinkurl{#1}}}

\bibitem{LIGOScientific:2016aoc}
B.~P. Abbott et~al. (LIGO Scientific, Virgo), Phys. Rev. Lett. \textbf{116},
  061102 (2016), arXiv: \eprint{1602.03837}.

\bibitem{LIGOScientific:2019fpa}
B.~P. Abbott et~al. (LIGO Scientific, Virgo), Phys. Rev. D \textbf{100}, 104036
  (2019), arXiv: \eprint{1903.04467}.

\bibitem{EventHorizonTelescope:2020qrl}
D.~Psaltis et~al. (Event Horizon Telescope), Phys. Rev. Lett. \textbf{125},
  141104 (2020), arXiv: \eprint{2010.01055}.

\bibitem{Penrose:1964wq}
R.~Penrose, Phys. Rev. Lett. \textbf{14}, 57 (1965).

\bibitem{ch5-}
S.~W. Hawking and G.~F.~R. Ellis,\textit{The Large Scale Structure of  Space-Time} (Cambridge University Press, Cambridge, 1973).

\bibitem{Witten:1991yr}
E.~Witten, Phys. Rev. D \textbf{44}, 314 (1991).

\bibitem{Rovelli:1997qj}
C.~Rovelli, in \textit{15th International Conference on General   Relativity and Gravitation (GR15)},  (Pune, India, 1997) pp. 281--331, arXiv:
  \eprint{gr-qc/9803024}.

\bibitem{Bojowald:2005qw}
M.~Bojowald, R.~Goswami, R.~Maartens, and P.~Singh, Phys. Rev. Lett.
  \textbf{95}, 091302 (2005), arXiv: \eprint{gr-qc/0503041}.

\bibitem{Saini:2014qpa}
A.~Saini and D.~Stojkovic, Phys. Rev. D \textbf{89}, 044003 (2014), arXiv:
  \eprint{1401.6182}.

\bibitem{Ashtekar:2021kfp}
A.~Ashtekar and E.~Bianchi, Rept. Prog. Phys. \textbf{84}, 042001 (2021),
  arXiv: \eprint{2104.04394}.

\bibitem{Barcelo:2015noa}
C.~Barcel\'o, R.~Carballo-Rubio, and L.~J. Garay, Universe \textbf{2}, 7
  (2016), arXiv: \eprint{1510.04957}.

\bibitem{Bardeen1968}
J.~Bardeen, in \textit{Proceedings of the 5th International Conference on  Gravitation and the Theory of Relativity},  (Tbilisi1968).

\bibitem{Hayward:2005gi}
S.~A. Hayward, Phys. Rev. Lett. \textbf{96}, 031103 (2006), arXiv:
  \eprint{gr-qc/0506126}.

\bibitem{Xiang:2013sza}
L.~Xiang, Y.~Ling, and Y.~G. Shen, Int. J. Mod. Phys. D \textbf{22}, 1342016
  (2013), arXiv: \eprint{1305.3851}.

\bibitem{Balart:2014jia}
L.~Balart and E.~C. Vagenas, Phys. Lett. B \textbf{730}, 14 (2014), arXiv:
  \eprint{1401.2136}.

\bibitem{Fan:2016hvf}
Z.-Y. Fan and X.~Wang, Phys. Rev. D \textbf{94}, 124027 (2016), arXiv:
  \eprint{1610.02636}.

\bibitem{Li:2016yfd}
X.~Li, Y.~Ling, Y.-G. Shen, C.-Z. Liu, H.-S. He, and L.-F. Xu, Annals Phys.
  \textbf{396}, 334 (2018), arXiv: \eprint{1611.09016}.

\bibitem{Nojiri:2017kex}
S.~Nojiri and S.~D. Odintsov, Phys. Rev. D \textbf{96}, 104008 (2017), arXiv:
  \eprint{1708.05226}.

\bibitem{He:2017ujy}
Y.~He and M.-S. Ma, Phys. Lett. B \textbf{774}, 229 (2017), arXiv:
  \eprint{1709.09473}.

\bibitem{Rodrigues:2018bdc}
M.~E. Rodrigues and M.~V. de~Sousa~Silva, JCAP \textbf{06}, 025 (2018), arXiv:
  \eprint{1802.05095}.

\bibitem{deSousaSilva:2018kkt}
M.~V. de~Sousa~Silva and M.~E. Rodrigues, Eur. Phys. J. C \textbf{78}, 638
  (2018), arXiv: \eprint{1808.05861}.

\bibitem{Simpson:2018tsi}
A.~Simpson and M.~Visser, JCAP \textbf{02}, 042 (2019), arXiv:
  \eprint{1812.07114}.

\bibitem{Zhang:2019acn}
C.~Zhang and X.~Zhang, Phys. Rev. D \textbf{101}, 086002 (2020), arXiv:
  \eprint{1912.07278}.

\bibitem{Melgarejo:2020mso}
G.~Melgarejo, E.~Contreras, and P.~Bargue\~no, Phys. Dark Univ. \textbf{30},
  100709 (2020).

\bibitem{Burzilla:2020utr}
N.~Burzill\`a, B.~L. Giacchini, T.~d.~P. Netto, and L.~Modesto, Eur. Phys. J. C
  \textbf{81}, 462 (2021), arXiv: \eprint{2012.11829}.

\bibitem{Bonanno:2020fgp}
A.~Bonanno, A.-P. Khosravi, and F.~Saueressig, Phys. Rev. D \textbf{103},
  124027 (2021), arXiv: \eprint{2010.04226}.

\bibitem{Estrada:2020tbz}
M.~Estrada and F.~Tello-Ortiz, EPL \textbf{135}, 20001 (2021), arXiv:
  \eprint{2012.05068}.

\bibitem{Torres:2022twv}
R.~Torres, {Regular Rotating Black Holes: A Review} (2022), arXiv:
  \eprint{2208.12713}.

\bibitem{Lan:2023cvz}
C.~Lan, H.~Yang, Y.~Guo, and Y.-G. Miao, Int. J. Theor. Phys. \textbf{62}, 202
  (2023), arXiv: \eprint{2303.11696}.

\bibitem{Junior:2023ixh}
J.~T. S.~S. Junior, F.~S.~N. Lobo, and M.~E. Rodrigues, Class. Quant. Grav.
  \textbf{41}, 055012 (2024), arXiv: \eprint{2310.19508}.

\bibitem{Mazza:2023iwv}
J.~Mazza and S.~Liberati, JHEP \textbf{03}, 199 (2023), arXiv:
  \eprint{2301.04697}.

\bibitem{Luongo:2023aib}
O.~Luongo and H.~Quevedo, Class. Quant. Grav. \textbf{41}, 125011 (2024),
  arXiv: \eprint{2305.11185}.

  \bibitem{Bianchi:2018mml}
E.~Bianchi, M.~Christodoulou, F.~D'Ambrosio, H.~M. Haggard, and C.~Rovelli,
  Class. Quant. Grav. \textbf{35}, 225003 (2018), arXiv: \eprint{1802.04264}.

\bibitem{DAmbrosio:2018wgv}
F.~D'Ambrosio and C.~Rovelli, Class. Quant. Grav. \textbf{35}, 215010 (2018),
  arXiv: \eprint{1803.05015}.


\bibitem{Rovelli:2018cbg}
C.~Rovelli and P.~Martin-Dussaud, Class. Quant. Grav. \textbf{35}, 147002
  (2018), arXiv: \eprint{1803.06330}.

\bibitem{Brahma:2018cgr}
S.~Brahma and D.-h. Yeom, Class. Quant. Grav. \textbf{35}, 205007 (2018),
  arXiv: \eprint{1804.02821}.

\bibitem{Carballo-Rubio:2018jzw}
R.~Carballo-Rubio, F.~Di~Filippo, S.~Liberati, and M.~Visser, Phys. Rev. D
  \textbf{98}, 124009 (2018), arXiv: \eprint{1809.08238}.

\bibitem{Bodendorfer:2019jay}
N.~Bodendorfer, F.~M. Mele, and J.~M\"unch, Class. Quant. Grav. \textbf{38},
  095002 (2021), arXiv: \eprint{1912.00774}.

\bibitem{Alesci:2019pbs}
E.~Alesci, S.~Bahrami, and D.~Pranzetti, Phys. Lett. B \textbf{797}, 134908
  (2019), arXiv: \eprint{1904.12412}.

\bibitem{BenAchour:2020gon}
J.~Ben~Achour, S.~Brahma, S.~Mukohyama, and J.~P. Uzan, JCAP \textbf{09}, 020
  (2020), arXiv: \eprint{2004.12977}.

\bibitem{Ling:2021olm}
Y.~Ling and M.-H. Wu, Class. Quant. Grav. \textbf{40}, 075009 (2023), arXiv:
  \eprint{2109.05974}.

\bibitem{LingLingYi:2021rfn}
Y.~Ling and M.-H. Wu, Chin. Phys. C \textbf{46}, 025102 (2022), arXiv:
  \eprint{2109.12938}.

\bibitem{Rignon-Bret:2021jch}
A.~Rignon-Bret and C.~Rovelli, Phys. Rev. D \textbf{105}, 086003 (2022), arXiv:
  \eprint{2108.12823}.

\bibitem{Ling:2022vrv}
Y.~Ling and M.-H. Wu, Symmetry \textbf{14}, 2415 (2022), arXiv:
  \eprint{2205.08919}.

\bibitem{Zeng:2022yrm}
W.~Zeng, Y.~Ling, and Q.-Q. Jiang, Chin. Phys. C \textbf{47}, 085103 (2023),
  arXiv: \eprint{2207.07529}.

\bibitem{Zeng:2023fqy}
W.~Zeng, Y.~Ling, Q.-Q. Jiang, and G.-P. Li, Phys. Rev. D \textbf{108}, 104072
  (2023), arXiv: \eprint{2308.00976}.

\bibitem{Feng:2023pfq}
Z.~Feng, Y.~Ling, X.~Wu, and Q.~Jiang, Sci. China Phys. Mech. Astron.
  \textbf{67}, 270412 (2024), arXiv: \eprint{2308.15689}.

\bibitem{Ayon-Beato:1998hmi}
E.~Ayon-Beato and A.~Garcia, Phys. Rev. Lett. \textbf{80}, 5056 (1998), arXiv:
  \eprint{gr-qc/9911046}.

\bibitem{Dehghani:2019xhm}
M.~Dehghani, Phys. Rev. D \textbf{99}, 104036 (2019).

\bibitem{Ayon-Beato:1999kuh}
E.~Ayon-Beato and A.~Garcia, Phys. Lett. B \textbf{464}, 25 (1999), arXiv:
  \eprint{hep-th/9911174}.

\bibitem{Bronnikov:2000vy}
K.~A. Bronnikov, Phys. Rev. D \textbf{63}, 044005 (2001), arXiv:
  \eprint{gr-qc/0006014}.

\bibitem{Dymnikova:2004zc}
I.~Dymnikova, Class. Quant. Grav. \textbf{21}, 4417 (2004), arXiv:
  \eprint{gr-qc/0407072}.

\bibitem{Balart:2014cga}
L.~Balart and E.~C. Vagenas, Phys. Rev. D \textbf{90}, 124045 (2014), arXiv:
  \eprint{1408.0306}.

\bibitem{Allahyari:2019jqz}
A.~Allahyari, M.~Khodadi, S.~Vagnozzi, and D.~F. Mota, JCAP \textbf{02}, 003
  (2020), arXiv: \eprint{1912.08231}.

\bibitem{Guerrero:2020uhn}
M.~Guerrero and D.~Rubiera-Garcia, Phys. Rev. D \textbf{102}, 024005 (2020),
  arXiv: \eprint{2005.08828}.

\bibitem{Kruglov:2021stm}
S.~I. Kruglov, Annals Phys. \textbf{428}, 168449 (2021), arXiv:
  \eprint{2104.08099}.

\bibitem{Kokoska:2021lrn}
D.~Koko\v{s}ka and M.~Ortaggio, Phys. Rev. D \textbf{104}, 124051 (2021),
  arXiv: \eprint{2109.08886}.

\bibitem{Ali:2022zox}
A.~Ali and K.~Saifullah, Eur. Phys. J. C \textbf{82}, 131 (2022).

\bibitem{Capozziello:2023vvr}
S.~Capozziello and G.~G.~L. Nashed, Class. Quant. Grav. \textbf{40}, 205023
  (2023), arXiv: \eprint{2309.08894}.

\bibitem{Moffat:2014aja}
J.~W. Moffat, Eur. Phys. J. C \textbf{75}, 175 (2015), arXiv:
  \eprint{1412.5424}.

\bibitem{Modesto:2016max}
L.~Modesto and L.~Rachwal, Finite conformal quantum gravity and nonsingular
  spacetimes (2016), arXiv: \eprint{1605.04173}.

\bibitem{Simpson:2019cer}
A.~Simpson, P.~Martin-Moruno, and M.~Visser, Class. Quant. Grav. \textbf{36},
  145007 (2019), arXiv: \eprint{1902.04232}.

\bibitem{Carballo-Rubio:2019fnb}
R.~Carballo-Rubio, F.~Di~Filippo, S.~Liberati, and M.~Visser, Phys. Rev. D
  \textbf{101}, 084047 (2020), arXiv: \eprint{1911.11200}.

\bibitem{Carballo-Rubio:2021wjq}
R.~Carballo-Rubio, F.~Di~Filippo, S.~Liberati, and M.~Visser, JHEP \textbf{02},
  122 (2022), arXiv: \eprint{2111.03113}.

\bibitem{Bakopoulos:2021liw}
A.~Bakopoulos, C.~Charmousis, and P.~Kanti, JCAP \textbf{05}, 022 (2022),
  arXiv: \eprint{2111.09857}.

\bibitem{Zeng:2021kyb}
D.-f. Zeng, Nucl. Phys. B \textbf{977}, 115722 (2022), arXiv:
  \eprint{2112.12531}.

\bibitem{Li:2023yyw}
Z.-C. Li and H.~Lu, Eur. Phys. J. C \textbf{83}, 755 (2023), arXiv:
  \eprint{2303.16924}.

\bibitem{Nian:2023xmr}
J.~Nian, Hawking radiation, entanglement entropy, and information paradox of
  kerr black holes (2023), arXiv: \eprint{2312.14287}.

\bibitem{Lin:2023jqz}
W.-C. Lin, D.-h. Yeom, and D.~Stojkovic, Spacetime surgery for black hole
  fireworks (2023), arXiv: \eprint{2310.15508}.

  \bibitem{Ashtekar:2018lag}
A.~Ashtekar, J.~Olmedo, and P.~Singh, Phys. Rev. Lett. \textbf{121}, 241301
  (2018), arXiv: \eprint{1806.00648}.

\bibitem{Ashtekar:2018cay}
A.~Ashtekar, J.~Olmedo, and P.~Singh, Phys. Rev. D \textbf{98}, 126003 (2018),
  arXiv: \eprint{1806.02406}.




\bibitem{Nicolai:2005mc}
H.~Nicolai, K.~Peeters, and M.~Zamaklar, Class. Quant. Grav. \textbf{22}, R193
  (2005), arXiv: \eprint{hep-th/0501114}.

\bibitem{Bojowald:2005cb}
M.~Bojowald and R.~Swiderski, Class. Quant. Grav. \textbf{23}, 2129 (2006),
  arXiv: \eprint{gr-qc/0511108}.

\bibitem{Cartin:2006yv}
D.~Cartin and G.~Khanna, Phys. Rev. D \textbf{73}, 104009 (2006), arXiv:
  \eprint{gr-qc/0602025}.

\bibitem{Ashtekar:2005qt}
A.~Ashtekar and M.~Bojowald, Class. Quant. Grav. \textbf{23}, 391 (2006),
  arXiv: \eprint{gr-qc/0509075}.

\bibitem{Modesto:2005zm}
L.~Modesto, Class. Quant. Grav. \textbf{23}, 5587 (2006), arXiv:
  \eprint{gr-qc/0509078}.

\bibitem{Boehmer:2007ket}
C.~G. Boehmer and K.~Vandersloot, Phys. Rev. D \textbf{76}, 104030 (2007),
  arXiv: \eprint{0709.2129}.

\bibitem{Chiou:2008nm}
D.-W. Chiou, Phys. Rev. D \textbf{78}, 064040 (2008), arXiv:
  \eprint{0807.0665}.

\bibitem{Brannlund:2008iw}
J.~Brannlund, S.~Kloster, and A.~DeBenedictis, Phys. Rev. D \textbf{79}, 084023
  (2009), arXiv: \eprint{0901.0010}.

\bibitem{Reuter:2012id}
M.~Reuter and F.~Saueressig, New J. Phys. \textbf{14}, 055022 (2012), arXiv:
  \eprint{1202.2274}.

\bibitem{Dadhich:2015ora}
N.~Dadhich, A.~Joe, and P.~Singh, Class. Quant. Grav. \textbf{32}, 185006
  (2015), arXiv: \eprint{1505.05727}.

\bibitem{Corichi:2015xia}
A.~Corichi and P.~Singh, Class. Quant. Grav. \textbf{33}, 055006 (2016), arXiv:
  \eprint{1506.08015}.

\bibitem{Olmedo:2017lvt}
J.~Olmedo, S.~Saini, and P.~Singh, Class. Quant. Grav. \textbf{34}, 225011
  (2017), arXiv: \eprint{1707.07333}.

\bibitem{Cortez:2017alh}
J.~Cortez, W.~Cuervo, H.~A. Morales-T\'ecotl, and J.~C. Ruelas, Phys. Rev. D
  \textbf{95}, 064041 (2017), arXiv: \eprint{1704.03362}.

\bibitem{Perez:2017cmj}
A.~Perez, Rept. Prog. Phys. \textbf{80}, 126901 (2017), arXiv:
  \eprint{1703.09149}.

\bibitem{Yonika:2017qgo}
A.~Yonika, G.~Khanna, and P.~Singh, Class. Quant. Grav. \textbf{35}, 045007
  (2018), arXiv: \eprint{1709.06331}.

\bibitem{Joe:2014tca}
A.~Joe and P.~Singh, Class. Quant. Grav. \textbf{32}, 015009 (2015), arXiv:
  \eprint{1407.2428}.

\bibitem{Afrin:2022ztr}
M.~Afrin, S.~Vagnozzi, and S.~G. Ghosh, Astrophys. J. \textbf{944}, 149 (2023),
  arXiv: \eprint{2209.12584}.

\bibitem{Martin-Dussaud:2019wqc}
P.~Martin-Dussaud and C.~Rovelli, Class. Quant. Grav. \textbf{36}, 245002
  (2019), arXiv: \eprint{1905.07251}.

\bibitem{Kelly:2020uwj}
J.~G. Kelly, R.~Santacruz, and E.~Wilson-Ewing, Phys. Rev. D \textbf{102},
  106024 (2020), arXiv: \eprint{2006.09302}.

\bibitem{Zhang:2020qxw}
C.~Zhang, Y.~Ma, S.~Song, and X.~Zhang, Phys. Rev. D \textbf{102}, 041502
  (2020), arXiv: \eprint{2006.08313}.

\bibitem{Zhang:2021wex}
C.~Zhang, Y.~Ma, S.~Song, and X.~Zhang, Phys. Rev. D \textbf{105}, 024069
  (2022), arXiv: \eprint{2107.10579}.

\bibitem{Zhang:2023yps}
X.~Zhang, Universe \textbf{9}, 313 (2023), arXiv: \eprint{2308.10184}.

\bibitem{Geiller:2020xze}
M.~Geiller, E.~R. Livine, and F.~Sartini, SciPost Phys. \textbf{10}, 022
  (2021), arXiv: \eprint{2010.07059}.

\bibitem{Daghigh:2020fmw}
R.~G. Daghigh, M.~D. Green, and G.~Kunstatter, Phys. Rev. D \textbf{103},
  084031 (2021), arXiv: \eprint{2012.13359}.

\bibitem{Ashtekar:2020ifw}
A.~Ashtekar, Universe \textbf{6}, 21 (2020), arXiv: \eprint{2001.08833}.

\bibitem{Hong:2022thd}
D.~K. Hong, W.-C. Lin, and D.-h. Yeom, Phys. Rev. D \textbf{106}, 104011
  (2022), arXiv: \eprint{2207.03183}.

\bibitem{ElizagaNavascues:2022npm}
B.~Elizaga~Navascu\'es, A.~Garc\'\i{}a-Quismondo, and G.~A. Mena~Marug\'an,
  Phys. Rev. D \textbf{106}, 063516 (2022), arXiv: \eprint{2207.04677}.

\bibitem{Lewandowski:2022zce}
J.~Lewandowski, Y.~Ma, J.~Yang, and C.~Zhang, Phys. Rev. Lett. \textbf{130},
  101501 (2023), arXiv: \eprint{2210.02253}.

\bibitem{Han:2023wxg}
M.~Han, C.~Rovelli, and F.~Soltani, Phys. Rev. D \textbf{107}, 064011 (2023),
  arXiv: \eprint{2302.03872}.

\bibitem{Zhang:2023okw}
C.~Zhang, Y.~Ma, and J.~Yang, Phys. Rev. D \textbf{108}, 104004 (2023), arXiv:
  \eprint{2302.02800}.

\bibitem{Fu:2023drp}
G.~Fu, D.~Zhang, P.~Liu, X.-M. Kuang, and J.-P. Wu, Phys. Rev. D \textbf{109},
  026010 (2024), arXiv: \eprint{2301.08421}.

\bibitem{Gan:2022oiy}
W.~C.~Gan, X.~M.~Kuang, Z.~H.~Yang, Y.~Gong, A.~Wang and B.~Wang, Sci. China Phys. Mech. Astron. \textbf{67}, 280411 (2024),
arXiv: \eprint{2212.14535}.

\bibitem{Gan:2024rga}
W.~C.~Gan and A.~Wang, Phys. Rev. D \textbf{111},   026017 (2025), arXiv: \eprint{2408.04494}.


\bibitem{Bojowald:2001xe}
M.~Bojowald, Phys. Rev. Lett. \textbf{86}, 5227 (2001), arXiv:
  \eprint{gr-qc/0102069}.

\bibitem{Ashtekar:2003hd}
A.~Ashtekar, M.~Bojowald, and J.~Lewandowski, Adv. Theor. Math. Phys.
  \textbf{7}, 233 (2003), arXiv: \eprint{gr-qc/0304074}.

\bibitem{Ashtekar:2004eh}
A.~Ashtekar and J.~Lewandowski, Class. Quant. Grav. \textbf{21}, R53 (2004),
  arXiv: \eprint{gr-qc/0404018}.

\bibitem{Thiemann:2006cf}
T.~Thiemann, Lect. Notes Phys. \textbf{721}, 185 (2007), arXiv:
  \eprint{hep-th/0608210}.

\bibitem{Thiemann:2001gmi}
T.~Thiemann, Modern canonical quantum general relativity (2001), arXiv:
  \eprint{gr-qc/0110034}.

\bibitem{Thiemann:2002nj}
T.~Thiemann, Lect. Notes Phys. \textbf{631}, 41 (2003), arXiv:
  \eprint{gr-qc/0210094}.

\bibitem{zhang2024blackholescovarianceeffective}
C.~Zhang, J.~Lewandowski, Y.~Ma, and J.~Yang, Black holes and covariance in
  effective quantum gravity (2024), arXiv: \eprint{2407.10168}.

\bibitem{Ashtekar:1987gu}
A.~Ashtekar, Phys. Rev. D \textbf{36}, 1587 (1987).

\bibitem{Ongole:2023pbs}
G.~Ongole, P.~Singh, and A.~Wang, Phys. Rev. D \textbf{109}, 026015 (2024),
  arXiv: \eprint{2311.10166}.

\bibitem{Rovelli:2014cta}
C.~Rovelli and F.~Vidotto, Int. J. Mod. Phys. D \textbf{23}, 1442026 (2014),
  arXiv: \eprint{1401.6562}.

\bibitem{Barrau:2014hda}
A.~Barrau and C.~Rovelli, Phys. Lett. B \textbf{739}, 405 (2014), arXiv:
  \eprint{1404.5821}.

\bibitem{DeLorenzo:2014pta}
T.~De~Lorenzo, C.~Pacilio, C.~Rovelli, and S.~Speziale, Gen. Rel. Grav.
  \textbf{47}, 41 (2015), arXiv: \eprint{1412.6015}.

\bibitem{Rovelli:2017zoa}
C.~Rovelli, Nature Astron. \textbf{1}, 0065 (2017), arXiv: \eprint{1708.01789}.

\bibitem{Lin:2024flv}
J.~Lin and X.~Zhang, Phys. Rev. D \textbf{110}, 026002 (2024), arXiv:
  \eprint{2402.13638}.

\bibitem{Nariai1951OnAN}
H.~Nariai, Gen. Rel. Grav. \textbf{31}, 963 (1951).

\bibitem{Alonso_Bardaji_2023}
A.~Alonso-Bardaji, D.~Brizuela, and R.~Vera, Phys. Rev. D \textbf{107}, 064067
  (2023), arXiv: \eprint{2302.10619}.

\bibitem{belfaqih2024blackholeseffectiveloop}
I.~H. Belfaqih, M.~Bojowald, S.~Brahma, and E.~I. Duque, Black holes in
  effective loop quantum gravity: Covariant holonomy modifications (2024),
  arXiv: \eprint{2407.12087}.

\bibitem{mato2024sphericallysymmetricloopquantum}
E.~Mato, J.~Olmedo, and S.~Saini, Spherically symmetric loop quantum gravity:
  Schwarzschild spacetimes with a cosmological constant (2024), arXiv:
  \eprint{2408.04925}.

\bibitem{Bojowald_2012}
M.~Bojowald and G.~M. Paily, Phys. Rev. D \textbf{86}, 104018 (2012), arXiv:
  \eprint{1112.1899}.

\bibitem{Alonso_Bardaji_2022}
A.~Alonso-Bardaji, D.~Brizuela, and R.~Vera, Phys. Lett. B \textbf{829}, 137075
  (2022), arXiv: \eprint{2112.12110}.

\bibitem{Alonso_Bardaji_2022_2}
A.~Alonso-Bardaji, D.~Brizuela, and R.~Vera, Phys. Rev. D \textbf{106}, 024035
  (2022), arXiv: \eprint{2205.02098}.

\bibitem{Bojowald_2024_1}
M.~Bojowald and E.~I. Duque, Phys. Rev. D \textbf{109}, 084006 (2024), arXiv:
  \eprint{2311.10693}.

\bibitem{Bojowald_2024_2}
M.~Bojowald, E.~I. Duque, and D.~Hartmann, Phys. Rev. D \textbf{109}, 084001
  (2024), arXiv: \eprint{2312.09217}.

\bibitem{Ashtekar_2020}
A.~Ashtekar and J.~Olmedo, Int. J. Mod. Phys. D \textbf{29}, 2050076 (2020),
  arXiv: \eprint{2005.02309}.

\end{thebibliography}
\end{document}